\documentclass[lettersize,journal]{IEEEtran}
\usepackage{amsmath,amsfonts}
\usepackage{algorithm}
\usepackage{array}
\usepackage[caption=false,font=scriptsize,labelfont=sf,textfont=sf]{subfig}
\usepackage{textcomp}
\usepackage{stfloats}
\usepackage{url}
\usepackage{verbatim}
\usepackage{graphicx}
\usepackage{cite}
\usepackage{pifont}

\newcommand{\fully}{\ding{52}}

\newcommand{\notso}{\ding{56}}
\usepackage[sort&compress,numbers]{natbib}
\newcolumntype{P}[1]{>{\centering\arraybackslash}p{#1}}
\usepackage{enumitem}
\setlist[itemize]{leftmargin=*, noitemsep}
\setlist[enumerate]{leftmargin=*, noitemsep}
\usepackage{xr}
\usepackage{amsmath,amsfonts}
\usepackage{graphicx}
\usepackage{textcomp}
\usepackage{xcolor}
\usepackage{comment}
\usepackage{booktabs}
\usepackage{enumitem}
\usepackage[caption=false]{subfig}
\usepackage{url}
\usepackage[normalem]{ulem}
\usepackage{algorithm}
\usepackage{algpseudocode}
\usepackage{array}
\usepackage{newtxmath}
\usepackage{euscript} 

\usepackage{amsmath}

\usepackage{colortbl}
\newcommand{\gray}{\cellcolor{gray!20}}

\begin{document}

\title{DASH: Deception-Augmented Shared Mental Model for a Human-Machine Teaming System}

\author{Zelin Wan, Han Jun Yoon, Nithin Alluru, Terrence J. Moore, Frederica F. Nelson, Seunghyun Yoon, Hyuk Lim, Dan Dongseong Kim, and Jin-Hee Cho,~\IEEEmembership{Senior Member, IEEE}
\IEEEcompsocitemizethanks{
\IEEEcompsocthanksitem This research is partially supported by the DEVCOM ARL Army Research Office (ARO) Award (W911NF-24-2-0241), the National Science Foundation (NSF) Secure and Trustworthy Cyberspace (SaTC) Award (2330940), and the Virtual Institutes for Cyber and Electromagnetic Spectrum Research and Employment (VICEROY) program under the Air Force Research Laboratory (AFRL) initiatives through The Griffiss Institute (419890). The views and conclusions contained in this document are those of the authors and should not be interpreted as representing the official policies, either expressed or implied, of the Army Research Laboratory or the U.S. Government. The U.S. Government is authorized to reproduce and distribute reprints for Government purposes, notwithstanding any copyright notation herein. ({\em Corresponding author: Zelin Wan}).  Zelin Wan, Han Jun Yoon, Nithin Alluru, and Jin-Hee Cho are with the Department of Computer Science, Virginia Tech, Arlington, VA, USA. Email: \{zelin, godzmdi93, nithin, jicho\}@vt.edu.  Terrence J. Moore and Frederica F. Nelson are with the US Army DEVCOM Army Research Laboratory, Adelphi, MD, USA. Email: \{terrence.j.moore.civ, frederica.f.nelson.civ\}@army.mil.  Seunghyun Yoon and Hyuk Lim are with the Korea Institute of Energy Technology (KENTECH), Naju-si, Jeollanam-do, Republic of Korea. Email: \{syoon, hlim\}@kentech.ac.kr.  Dan Dongseong Kim is with the University of Queensland, Brisbane, Queensland, Australia. Email: dan.kim@uq.edu.au.
}
}

\markboth{IEEE Transactions on Human-Machine Systems,~Vol.~xx, No.~x, xxx~2025}%
{Wan \MakeLowercase{\textit{et al.}}: DASH: Deception-Augmented Shared Mental Model for a Human-Machine Teaming System}


\maketitle

\begin{abstract}
We present DASH (Deception-Augmented Shared mental model for Human-machine teaming), a novel framework that enhances mission resilience by embedding proactive deception into Shared Mental Models (SMM). Designed for mission-critical applications such as surveillance and rescue, DASH introduces ``bait tasks'' to detect insider threats, e.g., compromised Unmanned Ground Vehicles (UGVs), AI agents, or human analysts, before they degrade team performance. Upon detection, tailored recovery mechanisms are activated, including UGV system reinstallation, AI model retraining, or human analyst replacement. In contrast to existing SMM approaches that neglect insider risks, DASH improves both coordination and security. Empirical evaluations across four schemes (DASH, SMM-only, no-SMM, and baseline) show that DASH sustains approximately 80\% mission success under high attack rates, eight times higher than the baseline. This work contributes a practical human-AI teaming framework grounded in shared mental models, a deception-based strategy for insider threat detection, and empirical evidence of enhanced robustness under adversarial conditions. DASH establishes a foundation for secure, adaptive human-machine teaming in contested environments.
\end{abstract}


\section{Introduction} \label{sec:introduction}

\textbf{Why human-machine teaming (HMT) systems?} As HMT systems are increasingly deployed in high-stakes operations, ensuring secure and efficient collaboration among human analysts, AI agents, and UGVs is critical~\cite{munasinghe2024comprehensive}. A well-established Shared Mental Model (SMM) enhances coordination by synchronizing tasks, fosters verification and validation, promotes adaptation to dynamic conditions, and mitigates failures. Without an effective SMM, miscommunication, inefficient task allocation, and increased risks of mission failure arise~\cite{bolstad1999shared}. Moreover, adversaries can exploit inconsistencies in trust and coordination, underscoring the need for a security-aware HMT framework that strengthens both collaboration and resilience.

\textbf{Why are SMMs critical?} According to Cannon-Bowers' team mental model framework~\cite{andrews2023role, cannon1993shared}, an SMM is known to establish a shared understanding that enhances coherent task execution and adaptability. However, their application in HMT systems remains underdeveloped~\cite{demir2020understanding}. Existing models~\cite{johnson2020understanding, sarker2023ai} fail to capture the dynamic interactions between human and AI teammates, and security remains largely unaddressed. Traditional static defenses~\cite{biggio2012poisoning} are insufficient against adversaries manipulating trust dynamics or selectively targeting system components. To ensure mission integrity, a security-aware SMM is essential for proactively detecting and mitigating these evolving threats.

\textbf{Cyber deception in HMT systems.}  Cyber deception has emerged as a promising proactive defense strategy~\cite{zhu2021survey}, offering unique advantages in securing HMT systems. Unlike conventional security mechanisms focusing solely on perimeter defense or reactive threat detection~\cite{vacca2013network, liao2013intrusion}, cyber deception actively manipulates an adversary’s perception, inducing suboptimal decisions~\cite{zhu2021survey}. This is particularly valuable in HMT environments, where the diverse attack surface, spanning human, AI, and physical components, creates complex trust relationships. By strategically deploying deceptive elements, such as bait tasks~\cite{bowen2009baiting}, defenders can identify compromised team members before mission degradation occurs~\cite{han2018CSUR}. The collaborative nature of HMT systems also enables cross-validation of actions across team members, further enhancing deception effectiveness~\cite{li2022unmanned, gay2019operator}. 

Moreover, deception techniques can seamlessly integrate with normal mission operations without disrupting workflow, preserving operational continuity~\cite{pawlick2019game}. Despite the critical need to secure HMT systems against sophisticated threats, particularly insider attacks that exploit the human-machine trust boundary, there is no existing comprehensive framework integrating cyber deception with HMT~\cite{andrews2023role, demir2020understanding}. This gap represents a significant vulnerability in mission-critical applications, where both performance and security must be jointly optimized~\cite{wang2022task}.

\textbf{Gaps filled by our approach, DASH.}  We propose \underline{D}eception-\underline{A}ugmented \underline{S}hared Mental Model for \underline{H}uman-Machine Teaming (DASH), integrating proactive cyber deception into SMMs to improve HMT performance and ensure security. Implemented in a simulated surveillance mission with UGVs, an AI agent, and a human analyst, DASH employs bait tasks for real-time insider threat detection and dynamic trust adjustment. It optimizes coordination through adaptive task allocation, balancing efficiency and security in adversarial environments. By formalizing an operational SMM framework, DASH integrates seamlessly into HMT systems, detecting compromised members without disrupting operations. It also evaluates trade-offs among mission success, operational costs, and detection effectiveness under varying APT conditions, ensuring resilient and secure human-machine collaboration.

This study makes the following \textbf{key contributions}:
\begin{itemize}
\item We integrate SMM with multiple UGVs in a simulated surveillance mission, enabling coordinated information sharing through Individual Mental Models (IMMs). This improves mission success rates by 25\% compared to systems without SMM, even under moderate attacks.

\item DASH combines SMM with cyber deception, using bait tasks for UGVs, AI agents, and human analysts. This proactive defense approach, with dynamic trust updates, detects compromised members early and significantly reduces compromise rates by up to 70\%.

\item We evaluate DASH across four schemes, assessing mission success, SMM quality, and resilience. DASH sustains 60\% mission success under high attack frequencies while ensuring cost-effectiveness against cyber threats.

\item Trust-based adaptation through the Adaptive Deceptive Task Management (ADTM) protocol enhances mission success and integrity. Security investments yield a 6 times improvement in mission success and significantly lower compromise rates, particularly for human analysts.

\end{itemize}
To our knowledge, no prior work has proposed a unified framework integrating SMMs with cyber deception, adaptive trust management, and cross-agent coordination to achieve robust and secure HMT under adversarial conditions.

\section{Background \& Related Work} \label{sec:related-work}

\subsection{Human-Machine Teaming (HMT) Systems}

\subsubsection{Definition} HMT refers to the collaboration between humans and machines to achieve shared objectives through interaction and communication~\cite{lyn2019opportunities}. \textit{Human-AI Teaming Systems (HATS)} integrate humans and AI models to combine their respective strengths~\cite{andrews2023role}, while \textit{Human-Machine Collaboration (HMC)} represents a collaborative form of Human-Machine Interface (HMI) common in health management, surveillance, and manufacturing~\cite{pizon2023human}. Our work implements an HMT system where humans and machines collaborate through coordinated information sharing and trust-based interactions.

\subsubsection{Trust} Trust in HMT systems encompass: (1) \textit{Trust in automation}, an agent's reliability in uncertain conditions~\cite{jiunyinin2000}, and (2) \textit{Mutual trust}, which enhances communication, consistency, and adaptability~\cite{ibrahim2022trust}. Machines must estimate human trust levels to ensure safety and interaction quality~\cite{alhaji2020toward}. Studies show how fidelity and trust influence human responses to autonomous systems~\cite{tossell2020appropriately}, with human oversight remaining critical in AI-driven operations~\cite{warren2020friend}. Our work implements trust as a dynamic component through the ADTM protocol, which adjusts information sharing based on trust levels to enhance mission integrity against insider threats.

\subsubsection{Shared Mental Models (SMMs)}
SMMs enhance coordination in HMT systems, improving team effectiveness~\cite{andrews2023role}. Computational models enable artificial agents to predict human behavior~\cite{scheutz2017framework,demir2020understanding}, while frameworks like CAST anticipate teammates' information needs~\cite{yen2003implementing,yen2006agents}. Robotic systems demonstrate that shared cognition enhances coordination in simulated environments~\cite{gervits2020toward}. Well-established SMMs improve communication and performance under stress~\cite{schelble2022let,stout1999planning,mathieu2000influence}, enhancing adaptability in dynamic environments~\cite{edgar2023improving,de2023managerial} across domains like surveillance, rescue, and logistics~\cite{carnegie2015affordable,wang2022task,ahn2015healthcare}.

Despite these advantages, existing SMM implementations have three critical limitations: they focus on coordination while neglecting security considerations; validation remains confined to simplified simulations (e.g., Minecraft-based USAR tests~\cite{demir2020understanding}); and they employ static trust assumptions~\cite{andrews2023role}. Our DASH framework addresses these through: (1) cyber deception mechanisms with bait tasks for detecting compromised agents, (2) dynamic trust management using multi-source verification, and (3) adaptive task allocation algorithms that maintain mission effectiveness during recovery cycles. DASH's validation demonstrates higher resilience to advanced persistent threat (APT) attacks compared to conventional approaches~\cite{mathieu2000influence, stout1999planning}.

\subsubsection{Mission-Driven HMT Systems}
Mission-driven HMT systems improve coordination in high-stakes settings~\cite{lematta2024practical}, with AI-as-Partner frameworks supporting complex task collaboration~\cite{mccomb2023focus}. Key advances include human systems engineering~\cite{boy2022machine}, model-based systems engineering for unmanned missions~\cite{michelson2019concepts}, resilient UGV platoon frameworks~\cite{li2022unmanned}, and hierarchical systems for disaster response~\cite{panagopoulos2022hierarchical}. Augmented reality enhances infantry, unmanned interaction~\cite{saleh2022exploring}, while psychological manipulation risks failing missions~\cite{steinmetz2022identification, wang2021social}.

However, existing systems lack real-time mechanisms for detecting compromised team members~\cite{mccomb2023focus,li2022unmanned}, sufficient integration between trust assessment and decision-making~\cite{gay2019operator,saleh2022exploring}, and adequate balance between security and mission performance~\cite{zweibelson2023part,lematta2024practical}. Studies on security threats~\cite{siddiqi2022study,abdelhamid2024power,femi2024systematic} highlight insider threats but lack integration with HMT frameworks. Our work addresses these limitations through proactive defense mechanisms that detect threats in real-time while maintaining operational continuity.

\subsection{Defensive Deception}
Defensive deception misleads attackers by shaping their perceptions to trigger suboptimal decisions~\cite{almeshekah2016cyber,rowe2016introduction}. Bait-based methods include trap files for unauthorized access detection~\cite{yuill2004honeyfiles}, game-theoretic honeypots~\cite{carroll2011game}, and systems combining camouflage and decoys~\cite{bowen2009baiting,keromytis2014systems}. Most existing techniques focus on conventional IT infrastructures with static trust models and are tested in isolated settings, not complex multi-agent systems.

DASH embeds bait tasks into task allocation and trust management to detect compromised team members while maintaining team performance. It is the first to integrate bait-based deception into an operational SMM for HMT systems, bridging traditional deception methods with the dynamic needs of modern human-machine teams.

\textbf{Table~\ref{tab:related_compare}} compares our work with existing HMT and deception frameworks based on their use of HMT, deception, SMM, and trust management. In Section~\ref{sec:result_and_analysis}, we further compare the performance of DASH against these counterparts, focusing on the impact of deception and SMM integration.

\begin{table*}[t]
  \centering
  \small
  \setlength{\tabcolsep}{4pt}
  \renewcommand{\arraystretch}{1.1}
  \caption{Comparison of HMT and Deception Frameworks on SMM, Deception, and Trust Dimensions}
  \label{tab:related_compare}
  \begin{tabular}{@{}P{2.9cm} c c c c p{10.4cm}@{}}
    \toprule
    \textbf{Work} & \textbf{HMT} & \textbf{Deception} & \textbf{SMM} & \textbf{Trust} & \multicolumn{1}{c}{\textbf{Key Technique}} \\
    \midrule
    \cite{jiunyinin2000} (2000) & \notso & \notso & \notso & \fully & Word elicitation, questionnaires, and factor analysis for measuring trust. \\
    \cite{yuill2004honeyfiles} (2004) & \notso & \fully & \notso & \notso & Honeyfiles triggering alerts on unauthorized access. \\
    \cite{almeshekah2016cyber} (2016) & \notso & \fully & \notso & \notso & Framework for planning honeypots and honeytokens. \\
    \cite{scheutz2017framework} (2017) & \fully & \notso & \fully & \notso & Framework for agents to build and synchronize shared mental models. \\
    \cite{demir2020understanding} (2020) & \fully & \notso & \fully & \notso & Discrete Recurrence Quantification of team coordination dynamics. \\
    \cite{panagopoulos2022hierarchical} (2022) & \fully & \notso & \notso & \notso & Hierarchical control switcher based on cognitive availability. \\
    \textbf{DASH (Ours)} & \fully & \fully & \fully & \fully & ADTM protocol enabling SMM, deception, and dynamic trust.\\
    \bottomrule
  \end{tabular}
  \\[4pt]
  {\footnotesize \fully: Considered; \notso: Not considered.}
\end{table*}

\section{Problem Statement}\label{sec:problem-statement}
This work considers a mission-oriented HMT system comprising multiple UGVs with peer-to-peer communication, an AI agent, a human analyst, and a command center for surveillance operations. Such systems are vulnerable to APTs that may implant backdoors in the AI agent, manipulate the human analyst via social engineering, or inject false data into UGV communications to disrupt mission goals. To counter these threats, the Command Center operates the DASH framework, managing trust, task allocation, and information integration. A Trusted Execution Environment (TEE)~\cite{munoz2023survey} ensures the integrity of critical computations and decisions.

Traditional security measures often fail to proactively detect or mitigate sophisticated threats~\cite{cho2020toward}. DASH addresses this gap by integrating individual mental models (IMMs) into a unified SMM and deploying deceptive ``bait tasks'' to identify compromised components. These tasks include planted objects for UGVs, validation prompts for the AI agent, and known-image checks for human analysts. A trust management mechanism continuously evaluates each node, triggering recovery protocols, UGV reinstallation, AI retraining, or analyst replacement, when trust scores fall below a threshold (see Section~\ref{subsec:trust-update-mechanisms}).

Each surveillance operation comprises $N_m$ missions, each representing a detection cycle (DC) initiated by the Command Center. A binary variable $C_i$ denotes success ($C_i = 1$ if classification is correct, $0$ otherwise). The total cost $C_{\text{total}}$ includes mission execution, bait deployment, and recovery.

The DASH framework aims to maximize:
\begin{equation}\label{eq:optimization_problem}
\arg\max \Big( \omega_1 \cdot \mathrm{MSR} \;-\; \omega_2 \cdot C_{\text{total}} \Big),
\end{equation}
where $\mathrm{MSR} = \frac{\sum_{i=1}^{N_m} C_i}{N_m}$, $\sum_{i=1}^{N_m} T_i \leq T_{\max}$, $T_i$ is the time for mission $i$ and $T_{\max}$ is the total time limit, and $\omega_1$ and $\omega_2$ ($\omega_1 + \omega_2 = 1$, $\omega_1, \omega_2 > 0$) balance mission success and cost.

As an illustrative example, consider a three-cycle mission where AI analysis achieves $\mathrm{MSR}=0.67$ at a total cost of $0.3$, while human review achieves $\mathrm{MSR}=1.00$ at a cost of $3.0$. With weights $\omega_1=0.7$ and $\omega_2=0.3$, \textbf{Eq.~\eqref{eq:optimization_problem}} yields a score of $0.7\times0.67 - 0.3\times0.3 \approx 0.38$ for AI analysis and $0.7\times1 - 0.3\times3 = -0.2$ for human review. Since the AI score is higher, AI analysis is preferred.

DASH integrates IMMs, an SMM, and cyber deception to proactively detect compromised members. It assigns regular and bait tasks to trigger recovery actions such as UGV reinstallation, AI retraining, or analyst replacement, ensuring mission resilience under adversarial threats (see Section~\ref{sec:dash}).

\section{System Model} \label{sec:system_model} 

\subsection{Network Model} \label{subsec:network_model}

The HMT system consists of multiple UGVs with autonomous navigation and image capture, communicating in a fully connected topology with the command center, the AI agent, and each other. UGVs share location, detection, and status data, enabling multi-angle verification to improve classification accuracy and mission reliability.

\begin{figure}
    \centering
    \includegraphics[width=\linewidth]{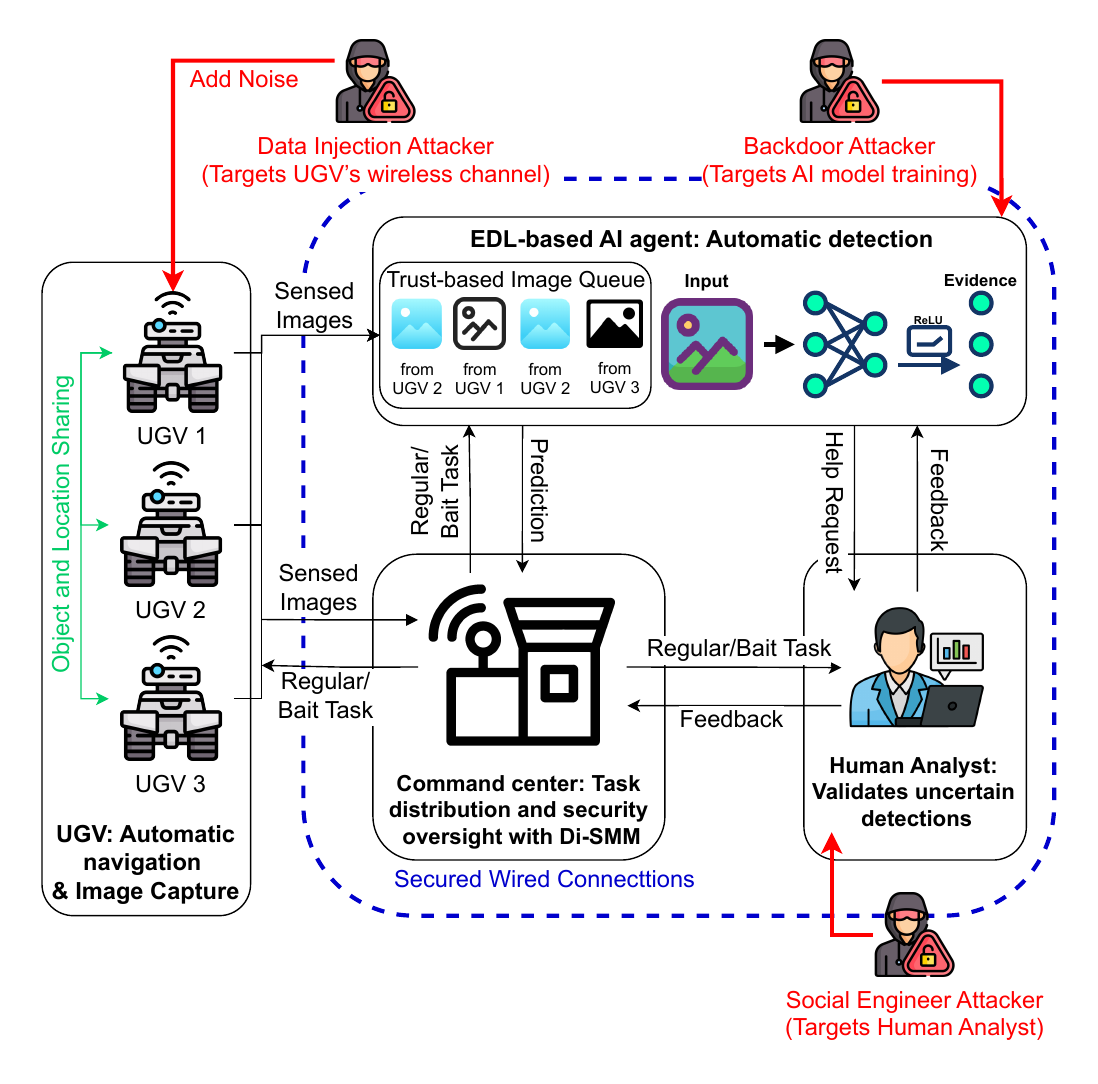}
    \caption{Overview of the proposed DASH-based HMT System.}
    \label{fig:human-AI-UGV-team}
\end{figure}

The AI agent processes detections and consults the human analyst under high uncertainty. The command center manages the mission, assigns regular and bait tasks, and monitors trust to maintain security. Wired links are secured via end-to-end encryption and physical controls~\cite{bameyi2021end}, while wireless channels remain vulnerable to injection and eavesdropping. This architecture enables efficient HMT with adaptive security, as illustrated in \textbf{Fig.~\ref{fig:human-AI-UGV-team}}.

\subsection{Node Model} \label{subsec:node_model}

\subsubsection{Command Center}
The command center initiates surveillance by specifying target object types and operates within a TEE to ensure secure task management. It implements the DASH framework by integrating IMMs into a unified SMM, allocating tasks, and deploying bait tasks to assess system integrity. Mission success is confirmed through cross-validation of AI and human analyst results.

\subsubsection{UGVs}
The system deploys autonomous UGVs with cameras and peer-to-peer communication, sharing location and detection data for multi-angle validation. Upon detecting a potential target, neighboring UGVs converge to confirm from different perspectives, relaying images to the AI agent and command center.

This validation process helps identify compromised UGVs injecting false data. If consensus is not achieved, the command center uses bait tasks to detect anomalies and triggers system reinstallation as needed.

\subsubsection{AI Agent}
The AI agent processes UGV images using a trust-based prioritized queue and applies Evidential Deep Learning (EDL)~\cite{sensoy2018evidential} to compute uncertain opinions via Subjective Logic (SL)~\cite{jsang2018subjective}. It generates a belief vector $\mathbf{b}$ and epistemic uncertainty $u$, ensuring $\sum \mathbf{b} + u = 1$. The per-detection cost is $C^A_{\mathrm{detect}} = 0.1$.

When $u > \tau_u$, the AI requests human input. The analyst provides a base rate vector $\mathbf{a}$ as prior class probabilities, incurring a higher cost $C^H_{\mathrm{detect}} = 1$. The final prediction is guided by $\mathbf{P}_x = \mathbf{b} + \mathbf{a} \cdot u$.

\subsubsection{Human Analyst}
The human analyst supports the system when AI uncertainty exceeds a threshold, contributing prior belief as a base rate vector $\mathbf{a} = [a_1, a_2, \ldots, a_C]$, with $\sum_{i=1}^{C} a_i = 1$~\cite{jsang2018subjective}. This belief is integrated into $\mathbf{P}_x = \mathbf{b} + \mathbf{a} \cdot u$ to enhance classification reliability.

\subsection{Threat Model} \label{subsec:threat_model}

The considered HMT system faces sophisticated adversarial threats targeting its key components. We categorize these threats as follows:

\begin{itemize}
    \item[(1)] \textbf{Compromising UGVs}: \textit{UGVs} are particularly vulnerable due to their reliance on wireless communication interfaces. They are susceptible to reconnaissance, false data injection, and command-and-control attacks that disrupt mission performance~\cite{atrews2020cyberwarfare, li2022unmanned}. Adversaries may analyze network traffic, inject false sensor data to mislead classification, or establish covert control channels to introduce transmission noise and impair coordination.
    
    \item[(2)] \textbf{Compromising AI agents}: \textit{AI agents} face backdoor attacks during training, where adversaries tamper with loss computation mechanisms to generate backdoored inputs~\cite{bagdasaryan2021blind}. These attacks are designed to preserve standard performance while embedding malicious behavior~\cite{desideri2012multiple}, and often remain undetected by activating near convergence using defense evasion strategies~\cite{wang2019neural}.
    
    \item[(3)] \textbf{Compromising human analysts}: \textit{Human analysts} are vulnerable to social engineering and psychological manipulation~\cite{steinmetz2022identification, wang2021social}. Adversaries may gain trust through seemingly credible interactions and exploit both psychological factors and material incentives~\cite{abdelhamid2024power, femi2024systematic}, potentially leading to biased or incorrect decision-making during high-stakes operations.
\end{itemize}
\textbf{Table~\ref{tab:ugv_attacks}} summarizes the attack behaviors and corresponding countermeasures proposed in this work. The DASH framework addresses these threats through a comprehensive, multi-layered defense as below.

\subsection{Defense Model} \label{subsec:defense-model}
To ensure mission resilience, DASH implements targeted defenses against compromised components in the HMT system. These countermeasures are designed to detect, isolate, and recover from adversarial disruptions while preserving operational continuity. They are described as follows.

\begin{itemize}
    \item \textbf{Defense Against Compromised UGVs:} DASH employs a hardware-based module to initiate secure remote reinstallation. Upon compromise detection, an encrypted command triggers system sanitization, and integrity is verified using cryptographic signatures.

    \item \textbf{Defense Against Compromised AI Agents:} DASH mitigates backdoor risks by verifying scripts, dependencies, and loss functions, followed by retraining AI models with trusted datasets and verified software.

    \item \textbf{Defense Against Compromised Human Analysts:} Compromised analysts are promptly replaced with standby personnel. DASH supports continuity through pre-trained alternates and reinforces resilience via security training and psychological support.

    \item \textbf{Defense for Operational Continuity:} DASH provisions backup resources, five standby UGVs, three AI agents, and alternate analysts, to maintain uninterrupted operations. These resources take over during recovery periods, with downtime durations of $\mathrm{DT}_{\mathrm{UGV}} = 2$, $\mathrm{DT}_{\mathrm{AI}} = 3$, and $\mathrm{DT}_{\mathrm{human}} = 5$.

\end{itemize}
\textbf{Table~\ref{tab:ugv_attacks}} summarizes the proposed countermeasures against the identified threats.

\begin{table*}[h]
    \centering
    \caption{Attack Behaviors and Countermeasures in DASH}
    \label{tab:ugv_attacks}
    \begin{tabular}{|P{3cm}|p{7cm}|p{7cm}|}
        \hline
        \textbf{Attack} & \multicolumn{1}{c|}{\textbf{Attack Behavior}} & \multicolumn{1}{c|}{\textbf{Countermeasures in DASH}} \\
        \hline
        \multicolumn{3}{|c|}{\bf \gray{UGV Attack Behaviors and Countermeasures}} \\
        \hline
        Reconnaissance~\cite{atrews2020cyberwarfare} & 
        Attackers analyze wireless traffic to identify vulnerabilities, monitor protocol patterns, and plan exploits. & 
        DASH detects anomalies, flags unusual scans, and dynamically adjusts security policies. \\
        \hline
        False Data Injection~\cite{li2022unmanned} & 
        Attackers manipulate sensor data to mislead the command center about UGV locations, detections, or status. & 
        DASH verifies data integrity via trust metrics and bait tasks, flagging anomalies for investigation. \\
        \hline
        C2-based Noise Injection~\cite{li2022unmanned} & 
        Attackers use a C2 channel to inject noise, degrading image quality or altering classifications. & 
        DASH applies multi-angle validation and human oversight to detect inconsistencies. \\
        \hline
        \multicolumn{3}{|c|}{\bf \gray{AI Agent Attack Behaviors and Countermeasures}} \\
        \hline
        Code-based Backdoor Injection~\cite{bagdasaryan2021blind} & 
        Attackers modify loss computation to create models that misclassify specific inputs while appearing normal. & 
        DASH audits training pipelines and libraries to detect and remove malicious code. \\
        \hline
        Defense Evasion~\cite{wang2019neural} & 
        Attackers embed evasion techniques that activate near convergence, improving resistance to detection. & 
        DASH issues bait tasks with known outputs to expose hidden model compromises. \\
        \hline
        Targeted Misclassification~\cite{desideri2012multiple} & 
        The compromised model misclassifies specific inputs while maintaining high normal accuracy. & 
        DASH retrains the AI model from scratch using verified datasets and trusted software. \\
        \hline
        \multicolumn{3}{|c|}{\bf \gray{Human Analyst Attack Behaviors and Countermeasures}} \\
        \hline
        Target Profiling~\cite{steinmetz2022identification} & 
        Attackers gather intelligence on analysts via social media and networks to exploit access. & 
        DASH cross-validates AI and human analyst outputs to detect anomalies. \\
        \hline
        Social Engineering~\cite{wang2021social, siddiqi2022study} & 
        Attackers craft pretexting scenarios to build credibility and escalate security-compromising requests. & 
        DASH deploys bait tasks to detect unusual response patterns indicative of external influence. \\
        \hline
        Psychological Manipulation~\cite{abdelhamid2024power, femi2024systematic} & 
        Attackers exploit incentives and psychological vulnerabilities to manipulate validation results. & 
        DASH triggers immediate personnel replacement, security training, and psychological support. \\
        \hline
    \end{tabular}
    \vspace{-3mm}
\end{table*}

\section{DASH Framework} \label{sec:dash}
This section presents DASH, a framework that optimizes task allocation, information sharing, and request handling in human-AI surveillance teams, while ensuring integrity through proactive threat detection. Building on traditional SMM theories~\cite{mathieu2000influence, stout1999planning, scheutz2017framework}, DASH integrates deception-based security to defend against adversarial threats.

DASH comprises three key components: individual mental models (IMMs), a shared mental model (SMM), and Adaptive Deceptive Task Management (ADTM) for secure task distribution. IMMs capture each participant's roles, tasks, and situational awareness, and are periodically merged into the SMM to support coordinated decision-making.

\subsection{Adaptive Deceptive Task Management (ADTM)} \label{subsec:adtm}
ADTM integrates deception into tasking by dynamically assigning regular and bait tasks to sustain mission continuity and detect compromised agents.

\begin{figure}[t]
    \centering
    \includegraphics[width=0.4\textwidth]{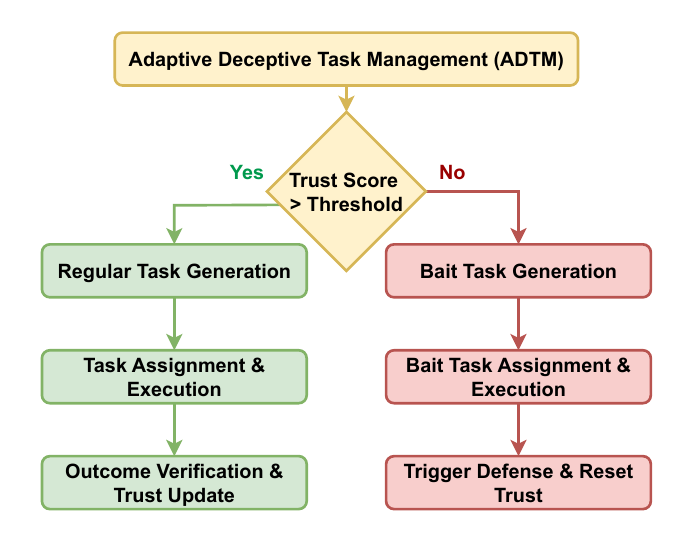}
    \vspace{-5mm}
    \caption{ADTM decision process for regular vs. bait task selection based on trust updates.}
    \label{fig:adtm_process}
    \vspace{-3mm}
\end{figure}

\begin{itemize}
    \item \textbf{Regular Task Creation and Distribution:} ADTM generates mission tasks (e.g., data collection, target identification, evidence analysis) using SMM data and assigns them dynamically based on roles, skills, and availability.

    \item \textbf{Bait Task-Based Detection Methods:} ADTM embeds bait tasks that mimic regular tasks but contain known outcomes visible only to the command center. Discrepancies in responses reduce the agent’s trust score $T_X$, and if $T_X < \zeta$ (e.g., 0.3), targeted integrity checks are triggered.

    \item \textbf{UGV Agent Detection:} Bait tasks instruct UGVs to capture images of strategically placed objects. Mismatched or missing data signals compromise and triggers low-impact defenses like secure system reinstallation.

    \item \textbf{AI Agent Detection:} AI agents receive bait tasks simulating false detections, cross-verified by the analyst. Inconsistencies lead to model retraining and source code inspection.

    \item \textbf{Human Analyst Detection:} Analysts are periodically tested with previously validated images. Incorrect classification indicates a possible compromise, prompting immediate personnel replacement to maintain system integrity.
\end{itemize}
Fig.~\ref{fig:adtm_process} illustrates the ADTM process, where regular task execution and trust updates trigger bait generation, outcome verification, and, if necessary, defense actions with trust resetting.

\subsection{Trust Management} \label{subsec:trust-update-mechanisms}
The system implements mechanisms to maintain member trustworthiness and operational integrity as follows.
\subsubsection{Chain-Based Trust Updates}  The command center updates trust scores by verifying task outcomes. When a UGV transmits an image, the AI agent’s classification is compared against the human analyst’s assessment. Agreement increases trust for all involved, while discrepancies reduce trust across the chain, signaling potential compromise. Pseudocode for the trust-score update process is provided as Algorithm~S1 in the supplement.
  
\subsubsection{Multi-UGV Mutual Verification}  To evaluate UGV trust accurately, the system uses multi-UGV validation. Each UGV starts with moderate trust (e.g., 0.5). When multiple UGVs detect the same object, the command center compares their outputs. Trust increases if they agree and are validated by the AI agent or human analyst. A deviating UGV is penalized; if all disagree, all lose trust.

\subsubsection{Trust Calculation Across Agents}  
Trust for member $X$ at time $t$ is calculated as:
\begin{equation} \label{eq:trust}
T_X(t) = \frac{N_C(t)+\lambda}{N_{\mathrm{total}}(t)+\lambda} \cdot e^{-\gamma t}, \quad T_X(t) \in [0,1],
\end{equation}
where $N_C(t)$ and $N_{\mathrm{total}}(t)$ are the number of consensus-matching and total tasks completed by $X$, respectively. The decay factor $\gamma = 0.001$ reduces trust over time, and $\lambda = 10$ ensures adaptive initialization.

If $T_X(t) < \zeta$, a bait task is issued. Success resets trust to 1; failure triggers defense protocols (Sec.~\ref{subsec:defense-model}). If trust remains below $\zeta$, the system initiates component-specific recovery: UGV reinstallation with standby units ($\mathrm{DT}_{\mathrm{UGV}} = 2$), AI retraining ($\mathrm{DT}_{\mathrm{AI}} = 3$), or analyst replacement ($\mathrm{DT}_{\mathrm{human}} = 5$). 

\subsubsection{Trust-Based Information Sharing} \label{subsubsec:adtm-info-sharing}
To enhance resilience, information-sharing frequency adapts to trust levels. For any interaction between members $X$ and $Y$, the probability of full data transmission is $T_Y(t)$, the trust of the recipient. High-trust members receive full data; low-trust ones get restricted access. This mechanism minimizes exposure of mission-critical information while incentivizing reliable behavior. Operating within the SMM, data sharing decreases as trust in $Y$ declines. Algorithm~S2 in the supplement formalizes this protocol.

\subsection{Strategic Shared Mental Model (SMM)}

\subsubsection{SMM Structure}
The SMM framework captures a team's collective understanding of system behavior, tasks, and roles, enhancing coordination and reducing miscommunication. It merges individual mental models (IMMs) into a unified reference that aligns members on objectives, execution, and responsibilities. Each team member maintains an IMM reflecting their unique knowledge, goals, state, and awareness, including mission plans, environmental knowledge, and expectations of others. While incomplete individually, IMMs form the foundation of shared understanding.

\subsubsection{Deception-Augmented Team Mental Model in the SMM} The SMM serves as the central hub of team cognition, continuously updated through information exchange. It supports role clarity, task coordination, mutual trust assessment, and real-time adaptation to situational changes. Deception-based verification mechanisms embedded within the SMM enable early detection of inconsistencies, which is critical for mission success (see Section~\ref{subsec:adtm}).

\begin{figure}[t]
    \centering
    \includegraphics[width=0.5\textwidth]{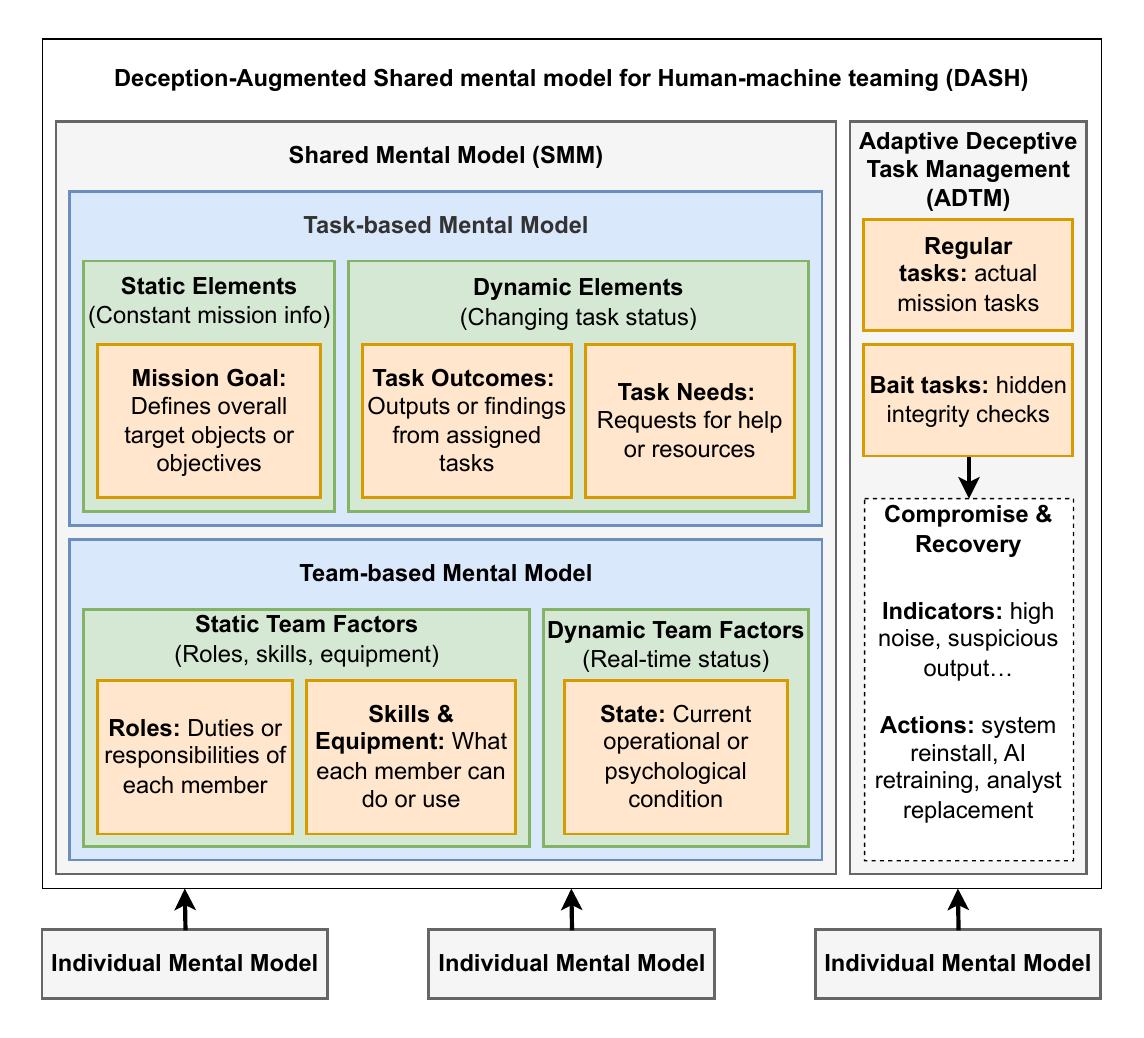}
    \vspace{-7mm}
    \caption{Our Proposed Shared Mental Model (SMM): A detailed SMM with extended explanations is in the supplementary document. 
    }
    \label{fig:smm_framework_small}
    \vspace{-3mm}
\end{figure}

Each bait task outcome updates the Team Mental Model (TEMM) component of the SMM by adjusting the trust score $T_X(t)$. These updated trust values inform team member actions; for example, the AI agent prioritizes its processing queue based on descending trust scores, favoring high-trust sources. Compact versions of the SMM and IMM frameworks are shown in Figs.~\ref{fig:smm_framework_small} and~\ref{fig:imm_framework_small}, respectively, with full diagrams provided in Figs.~S1--S2 of the supplement. A summary of key DASH components appears in \textbf{Table~SI} of the supplement document.

\subsubsection{Example SMM}  
\begin{figure}
    \centering
    \includegraphics[width=0.48\textwidth]{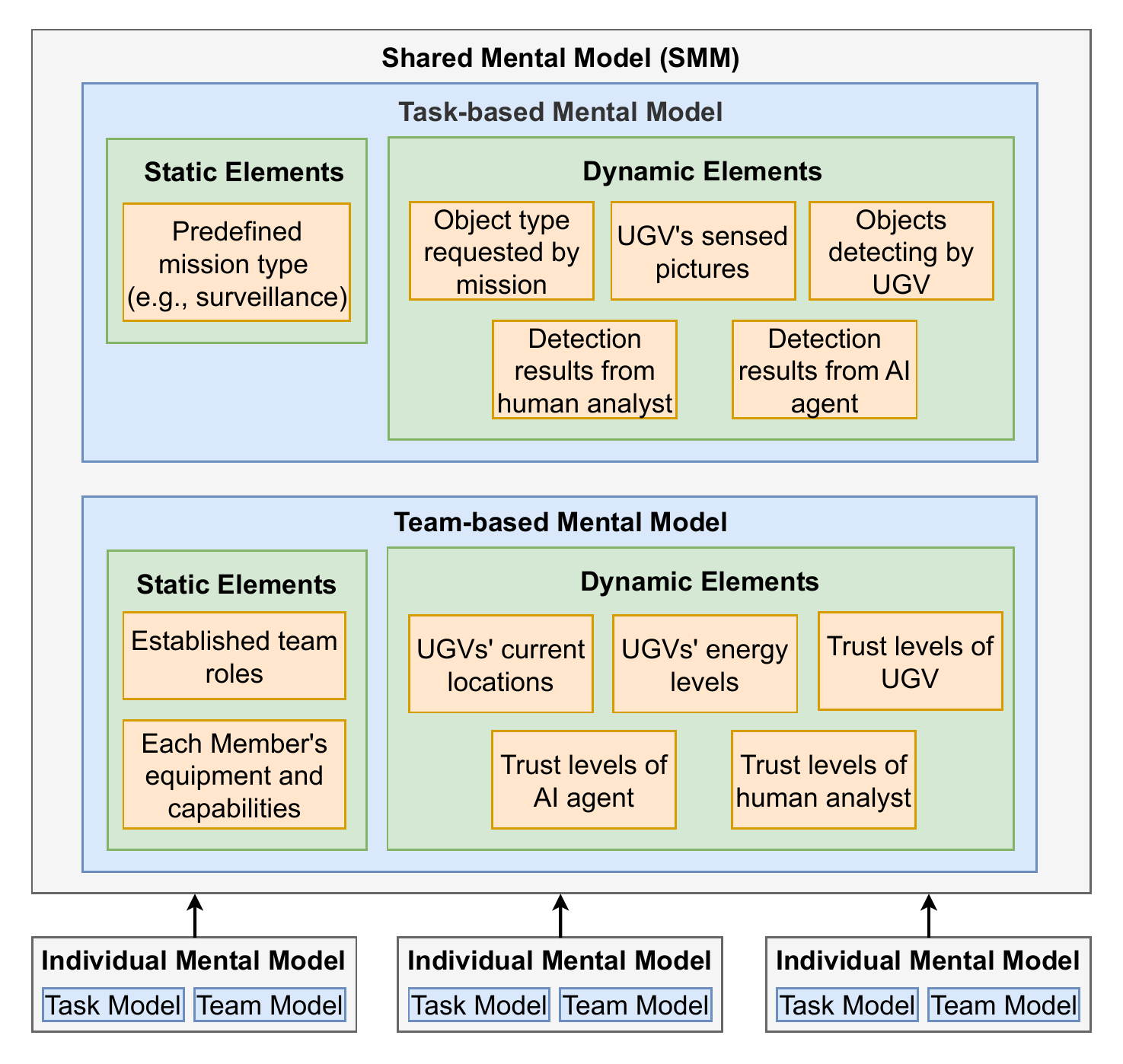}
    \vspace{-3mm}
    \caption{SMM example for the surveillance scenario.}
    \label{fig:smm-example}
    \vspace{-3mm}
\end{figure}

To illustrate, \textbf{Fig.~\ref{fig:smm-example}} shows an SMM for a surveillance mission involving a command center, UGVs, an AI agent, and a human analyst. Each team member maintains an IMM composed of a Task Mental Model (TAMM, $\mathrm{Task}_{MM}^I$) and a Team Mental Model (TEMM, $\mathrm{Team}_{MM}^I$). The TAMM includes static information (e.g., mission type, area maps) and dynamic elements (e.g., detected objects, sensor data, classification results), updated throughout the mission. The TEMM captures static factors (e.g., roles, capabilities) and dynamic updates (e.g., member locations, energy levels, operational status, trust).

The collective SMM integrates these IMMs. The shared TAMM ($\mathrm{Task}_{MM}^S$) aggregates static mission info and real-time updates on detections, classifications, and verifications. The shared TEMM ($\mathrm{Team}_{MM}^S$) consolidates team capability data with status updates on positions, queues, availability, and trust metrics derived from performance history.  This SMM framework ensures synchronized team understanding, improving coordination and mission effectiveness.

We summarize the mathematical notation used in our modeling and analysis in \textbf{Table~SII} of the supplementary document.

\begin{figure}[t]
    \centering
    \includegraphics[width=0.48\textwidth]{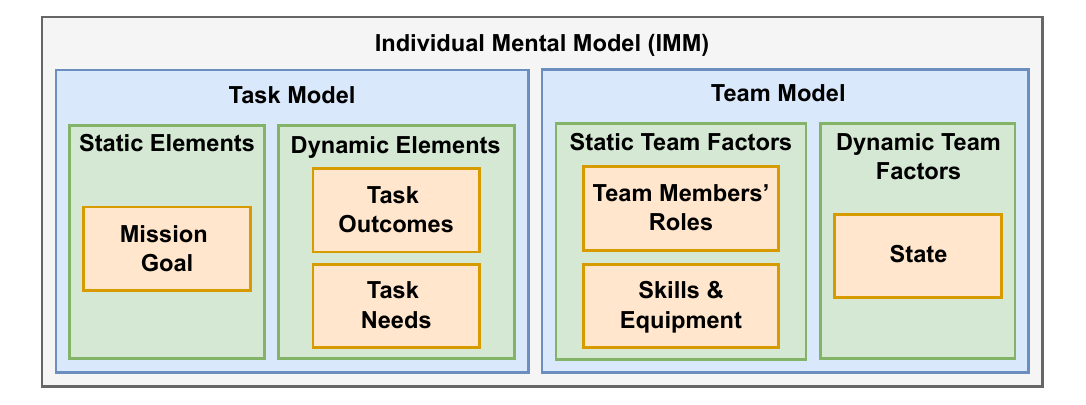}
    \vspace{-3mm}
\caption{Our Proposed Individual Mental Model (IMM): A detailed IMM with extended explanations is in the supplementary document.}
    \label{fig:imm_framework_small}
    \vspace{-3mm}
\end{figure}

\section{Experimental Setup} \label{sec:experiment_setup}

\subsection{Parameterization}

\subsubsection{Attack and Vulnerability Setting}
To evaluate DASH’s effectiveness, we simulate adversarial conditions where attack attempts occur equally often but exploit distinct vulnerabilities. UGVs are most vulnerable (30\% success rate) due to exposed wireless channels. AI agents face a 10\% risk from backdoor injections, mitigated by code audits. Human analysts, protected through security training, are least vulnerable (5\% success). These rates reflect real-world disparities: UGVs have broad attack surfaces, AI benefits from verifiability, and humans gain resilience through training.

\subsubsection{Detection Cycles}
A detection cycle is DASH’s core operational unit, initiated by the Command Center. Each cycle includes task dispatch, UGV sensing, AI/human analysis, and ends with object classification confirmation. 

\subsubsection{Mission}
A mission comprises repeated detection cycles targeting a specific class (e.g., 10 suspicious vehicles). (1) The Command Center assigns tasks to UGVs, each covering a designated area. (2) When a UGV captures a potential target, it uploads the image for AI analysis. If uncertainty is high, a human analyst is consulted (see Section~\ref{sec:system_model}). (3) Upon confirmation, other UGVs may be redirected for multi-angle validation or prioritized instance collection. Lower-priority images may be dropped to conserve processing. (4) The AI agent classifies the image, invoking human validation if uncertainty exceeds a threshold. (5) This process repeats until the required detections are reached. A mission succeeds if targets are identified within the time limit; otherwise, it fails.

\subsubsection{Environment} The environment includes five object types (i.e., car, truck, human, animal, and tree) for detection. We fine-tuned a ViT-B/16 vision transformer (pre-trained on ImageNet) using 6,000 CIFAR-10 surveillance images with balanced category representation. CIFAR-10 was selected for its publicly available, balanced object classes that align with our surveillance targets and support reproducible evaluation. The ViT-B/16 model was fine-tuned directly on the original CIFAR-10 images without additional preprocessing, focusing on demonstrating the capabilities of our EDL-based AI agent.

For Evidential Deep Learning (EDL)~\cite{sensoy2018evidential}, we modified the ViT output to produce evidence values for uncertainty quantification using Subjective Logic, enabling the system to determine when escalation to human input was warranted. Training was performed using the Adam optimizer with a learning rate of 0.0001, a batch size of 32, and 10,000 epochs.

\subsubsection{Simulation Setting}
\textbf{Table~SIII} in the supplement summarizes key simulation parameters chosen to balance realism and feasibility. The trust threshold ($\zeta = 0.3$) triggered bait tasks in response to suspected compromises, while the uncertainty threshold ($\tau_u = 0.25$) determined when the AI agent requested human validation. To ensure statistical significance, 100 simulations were conducted per scheme. Attack frequencies ranged from 0.0 to 1.0 in increments of 0.1, enabling evaluation across scenarios from benign to fully adversarial conditions.

\subsection{Metrics}
\begin{itemize}
\item \textbf{Mission Success Rate (MSR)} is the proportion of successfully completed missions:
\begin{equation} \label{eq:msr}
\text{MSR} = \frac{\sum_{i=1}^{N_m} C_i}{N_m}.
\end{equation}

\item \textbf{Ratio of Compromised Members} is the probability that a team member $X$ (UGV, AI, or human) remains compromised during a mission:
\begin{equation}
R_X = \frac{1}{N_{\text{sim}}} \sum_{i=1}^{N_{\text{sim}}} X_{\mathrm{comp}}^{(i)},
\end{equation}
where $X_{\mathrm{comp}}^{(i)} = 1$ if $X$ was compromised in mission $i$, 0 otherwise.

\item \textbf{Operational Cost (OC)} is the average cost per mission:
\begin{equation}
\text{OC} = \frac{C_{\text{total}}}{N_{\text{total}}},
\end{equation}
with
\begin{equation}
C_{\text{total}} = \sum_{i=1}^{N_{\text{total}}} (C_{\text{UGV}}^{(i)} + C_{\text{AI}}^{(i)} + C_{\text{human}}^{(i)} + C_{\text{recovery}}^{(i)}),
\end{equation}
where each term denotes mission-specific costs for UGVs, AI, human analysts, and recovery (e.g., reinstallation or retraining).

\item \textbf{SMM Quality Index (SQI)} measures the accuracy of information exchanged via the SMM:
\begin{equation}
\text{SQI} = \frac{|I_{\mathrm{correct}}|}{|I_{\mathrm{shared}}|},
\end{equation}
where $|I_{\mathrm{correct}}|$ is the number of correctly shared pieces, and $|I_{\mathrm{shared}}|$ is the total exchanged.

\item \textbf{SMM Coverage Index (SCI)} quantifies the efficiency of team information flow:
\begin{equation}
\text{SCI} = \frac{|I_{\mathrm{shared}}|}{|I_{\mathrm{max}}|},
\end{equation}
where $|I_{\mathrm{max}}|$ is the total possible information that could be shared.
\end{itemize}

The relationships between $I_{\mathrm{correct}}$, $I_{\mathrm{shared}}$, and $I_{\mathrm{potential}}$ are: $I_{\mathrm{correct}} \subseteq I_{\mathrm{shared}} \subseteq I_{\mathrm{max}}$.

\subsection{Comparing Schemes} \label{subsec:comparing-schemes}

To evaluate the DASH framework, we designed four experimental schemes with varying integration levels:

\begin{itemize}
    \item \textbf{DASH-DF (Deception-Augmented SMM with Defense):} Full implementation with SMM-based coordination, trust-based task allocation, and deception. UGVs share data for multi-angle verification, the AI agent prioritizes tasks by trust, and bait tasks confirm suspected compromises before defenses are triggered (see Section~\ref{subsec:defense-model}).

    \item \textbf{SMM-DF (SMM with Defense):} Same as DASH-DF but without deception; defenses trigger immediately when trust falls below threshold $\zeta$, without bait task verification.

    \item \textbf{DF-only (Defense-only):} No SMM coordination or UGV information sharing. The AI agent does not seek human input for uncertain cases. The command center updates trust, and defenses activate when trust falls below $\zeta$.

    \item \textbf{BASE (Baseline):} Conventional setup with standard task dispatch. UGVs, an AI agent, and a human analyst operate independently, with no trust updates or defense mechanisms, even when outputs conflict.
\end{itemize}
This setting systematically evaluates SMM and deception-based defenses on system performance and resilience.

\section{Simulation Results and Analyses} \label{sec:result_and_analysis}


\subsection{Comparative Performance Analysis} \label{sec:comparative_performance_analysis}
We evaluate the four schemes over 200 object detection (OD) missions. In each figure, the x-axis denotes OD missions (1–200), and the y-axis shows the corresponding metric. The results demonstrate the advantages of integrating SMMs and deception-based defenses.

\subsubsection{Mission Success Rate (MSR)} \label{sec:msr}
\textbf{Fig.~\ref{fig:msr_missions}} shows MSR trends over 200 missions. All schemes initially declined under adversarial pressure, stabilizing as defenses were activated. DASH-DF consistently achieved the highest MSR due to SMM coordination and deception-based detection, which reduced compromises and improved information flow. BASE performed the worst, lacking both mechanisms.

\subsubsection{SMM Quality Index (SQI)} \label{sec:smm_quality_index}
\textbf{Fig.~\ref{fig:smm_quality_index}} compares the SQI of all considered schemes. DASH-DF outperformed SMM-DF, as deception preserved data integrity and ensured accurate exchange. DF-only and BASE had zero SQI due to the absence of SMM, trust-sharing, UGV positioning, or AI-human collaboration, highlighting the importance of SMM in reliable communication.

\subsubsection{SMM Coverage Index (SCI)} \label{sec:smm_coverage_index}
\textbf{Fig.~\ref{fig:smm_coverage_index}} presents the SCI of all considered schemes. DASH-DF slightly outperformed SMM-DF. Unlike SQI, SCI reflects the completeness of information sharing rather than correctness. DF-only and BASE again scored zero, lacking inter-agent communication.

\subsubsection{Operational Cost} \label{sec:operational_cost}
\begin{figure*}[t]
\centering
\subfloat{\includegraphics[width=0.5\textwidth]{figures/metrics_legend.png}}
\hfil

\vspace{-3mm}
\setcounter{subfigure}{0}
\subfloat[UGV Operational Cost]{\includegraphics[width=0.25\textwidth]{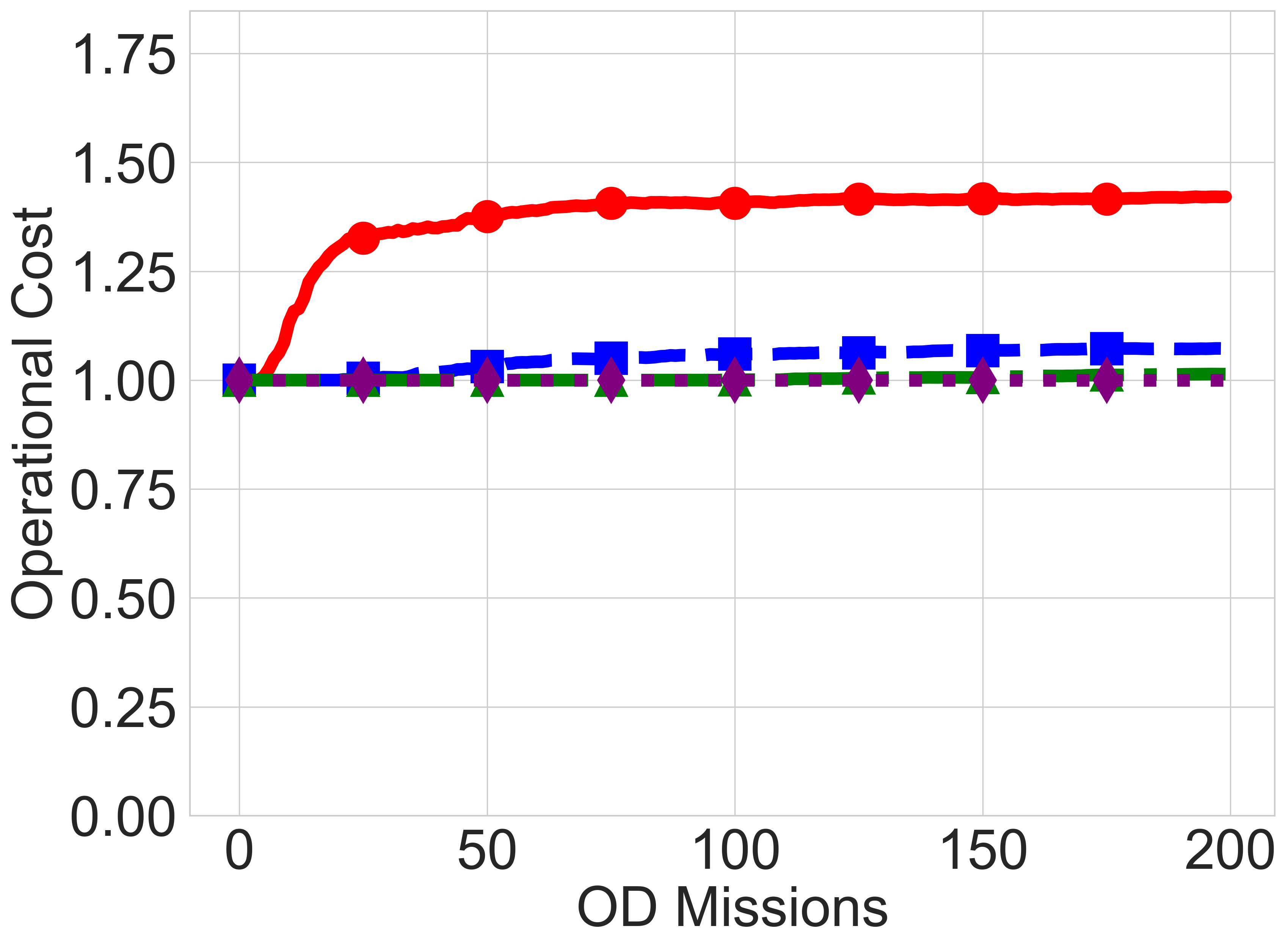}}
\hfil
\subfloat[AI Operational Cost]{\includegraphics[width=0.25\textwidth]{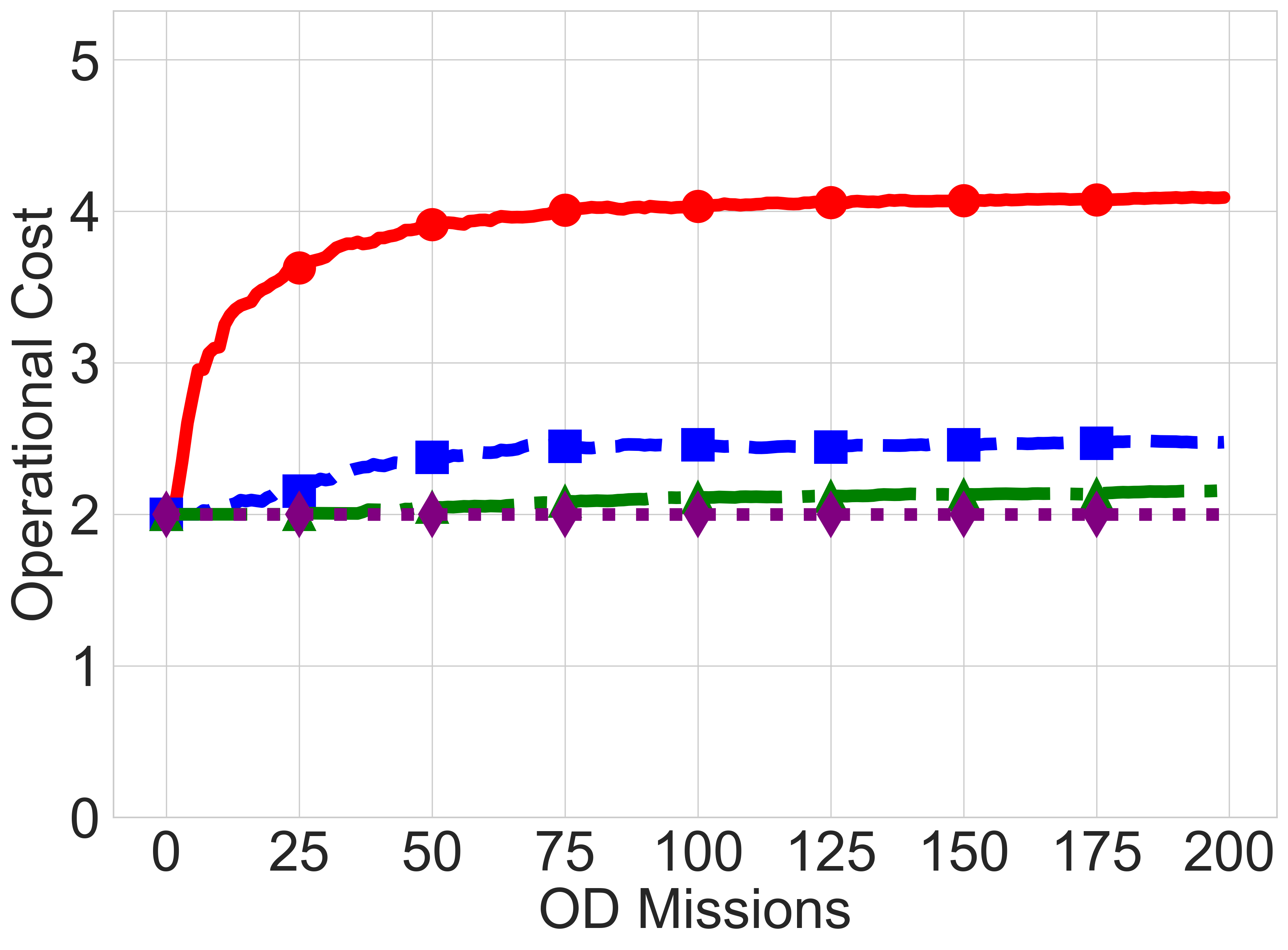}}
\hfil
\subfloat[Human Analysis Operational Cost]{\includegraphics[width=0.25\textwidth]{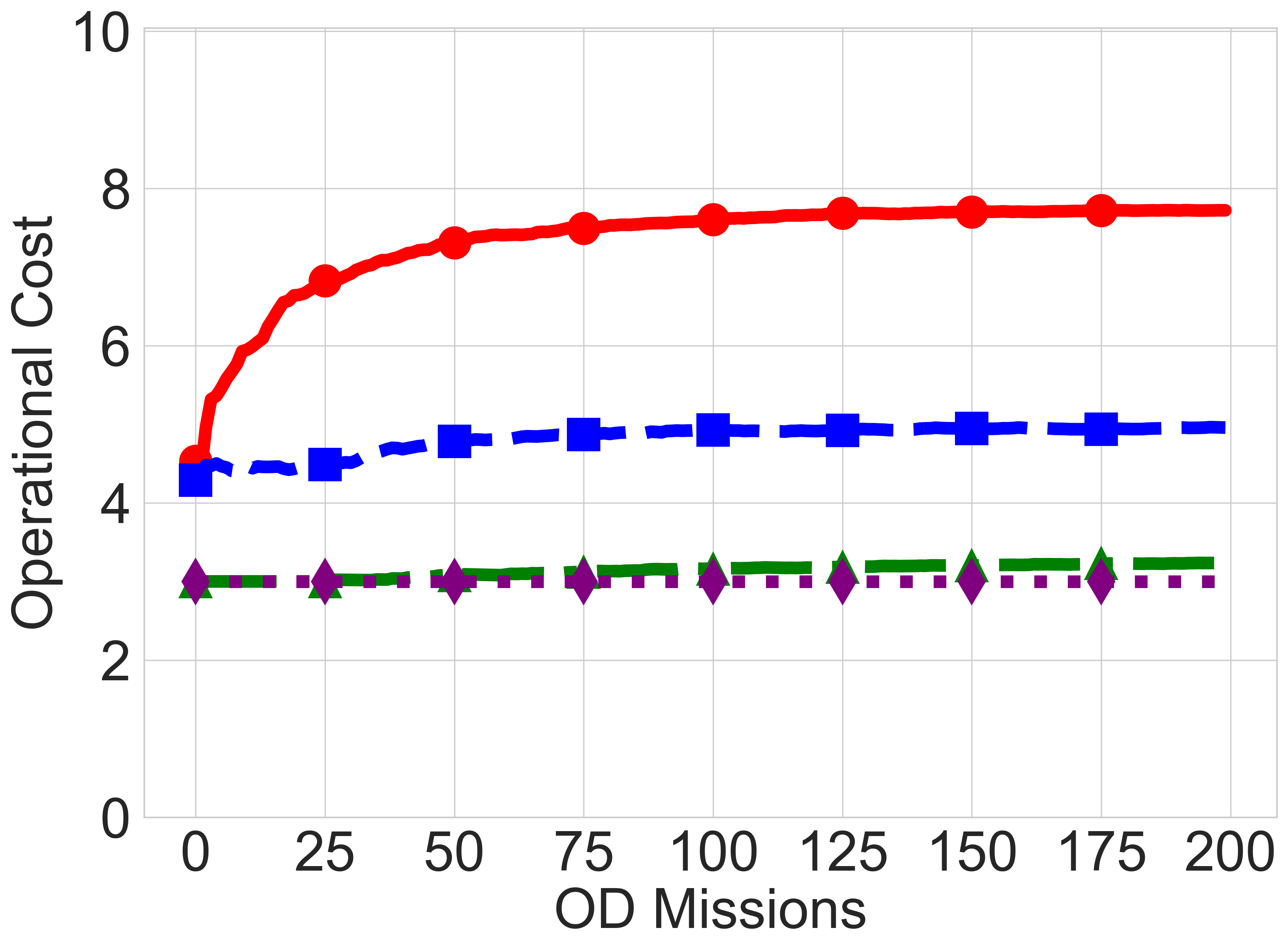}}
\hfil
\subfloat[Total Operational Cost of All Members]{\includegraphics[width=0.25\textwidth]{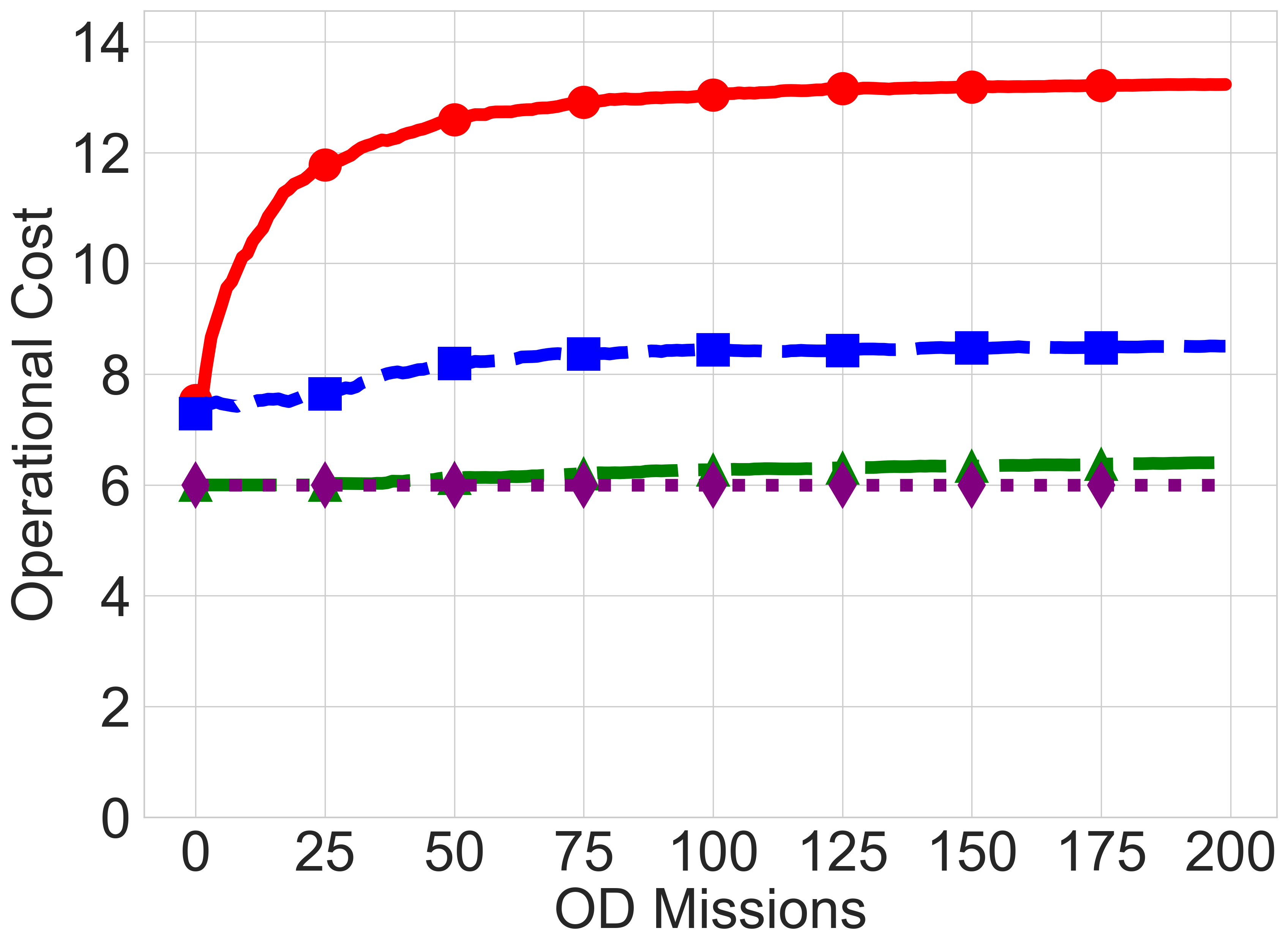}}
\hfil
\caption{Comparison of operational costs for four schemes under a fixed attack rate of 0.4.}
\label{fig:operational_cost}
\vspace{-3mm}
\end{figure*}

\textbf{Fig.~\ref{fig:operational_cost}} compares operational costs. DASH-DF incurred higher costs due to the strategic deployment of deception tasks based on dynamic trust. Human analyst verification was most expensive, as analyst-specific bait tasks required greater resources. Despite the cost, DASH-DF’s proactive defenses significantly reduced compromises (\textbf{Fig.~\ref{fig:compromised_members}}), justifying the investment in security.

\subsubsection{Ratio of Compromised Members} \label{sec:compromised_members}
\begin{figure*}[t]
\centering
\subfloat{\includegraphics[width=0.5\textwidth]{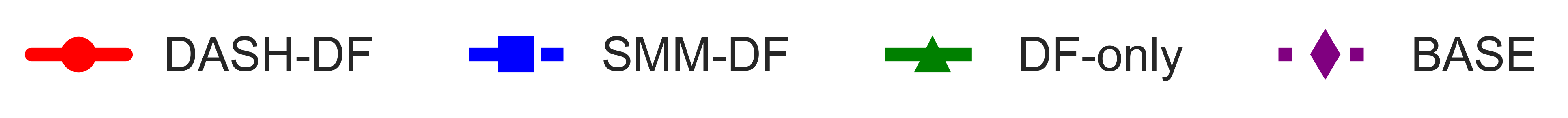}}
\hfil

\vspace{-3mm}
\setcounter{subfigure}{0}
\subfloat[Mission Success Rate ($\text{MSR}$)]{\includegraphics[width=0.25\textwidth]{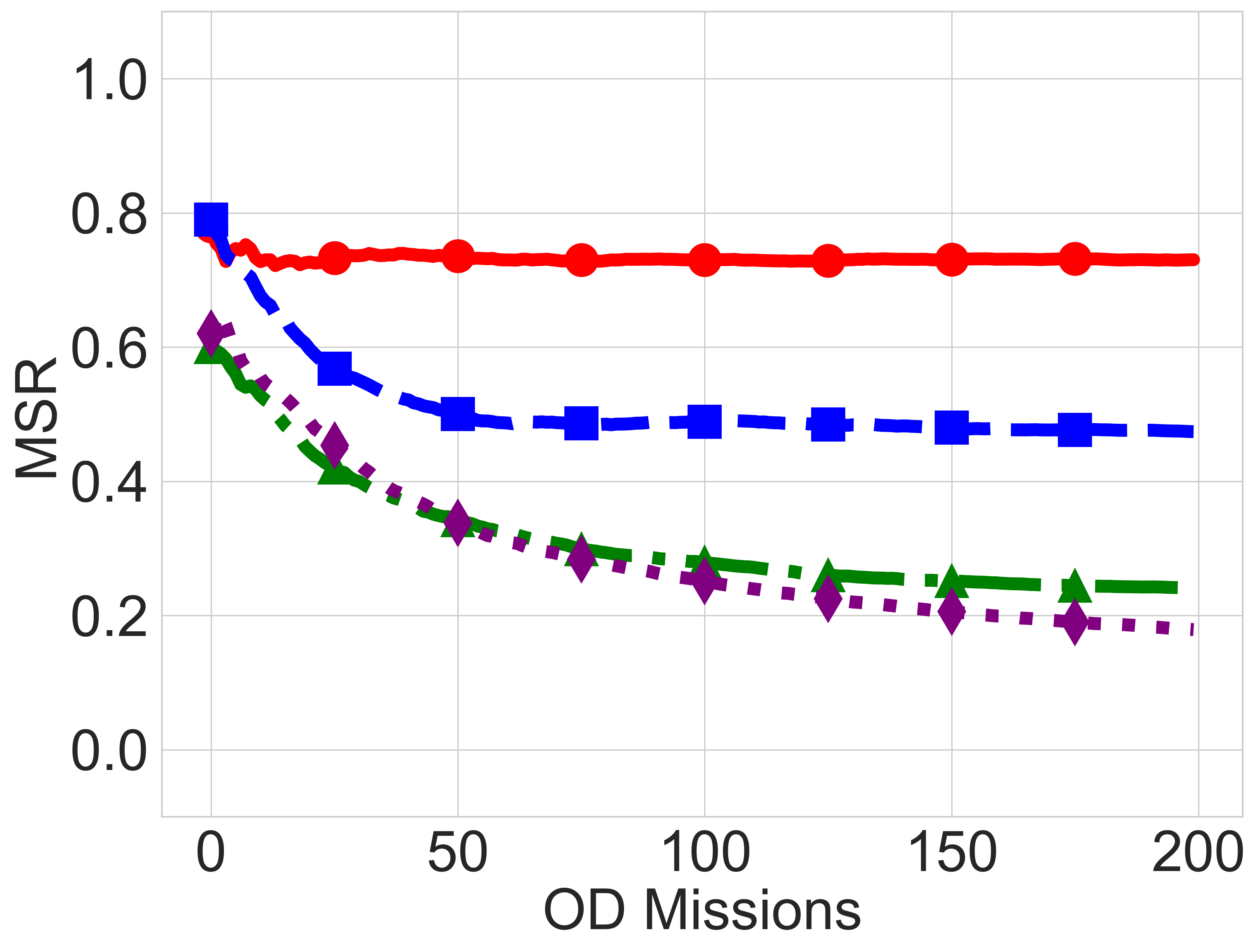}\label{fig:msr_missions}}
\hfil
\subfloat[SMM Quality Index ($\text{SQI}$)]{\includegraphics[width=0.25\textwidth]{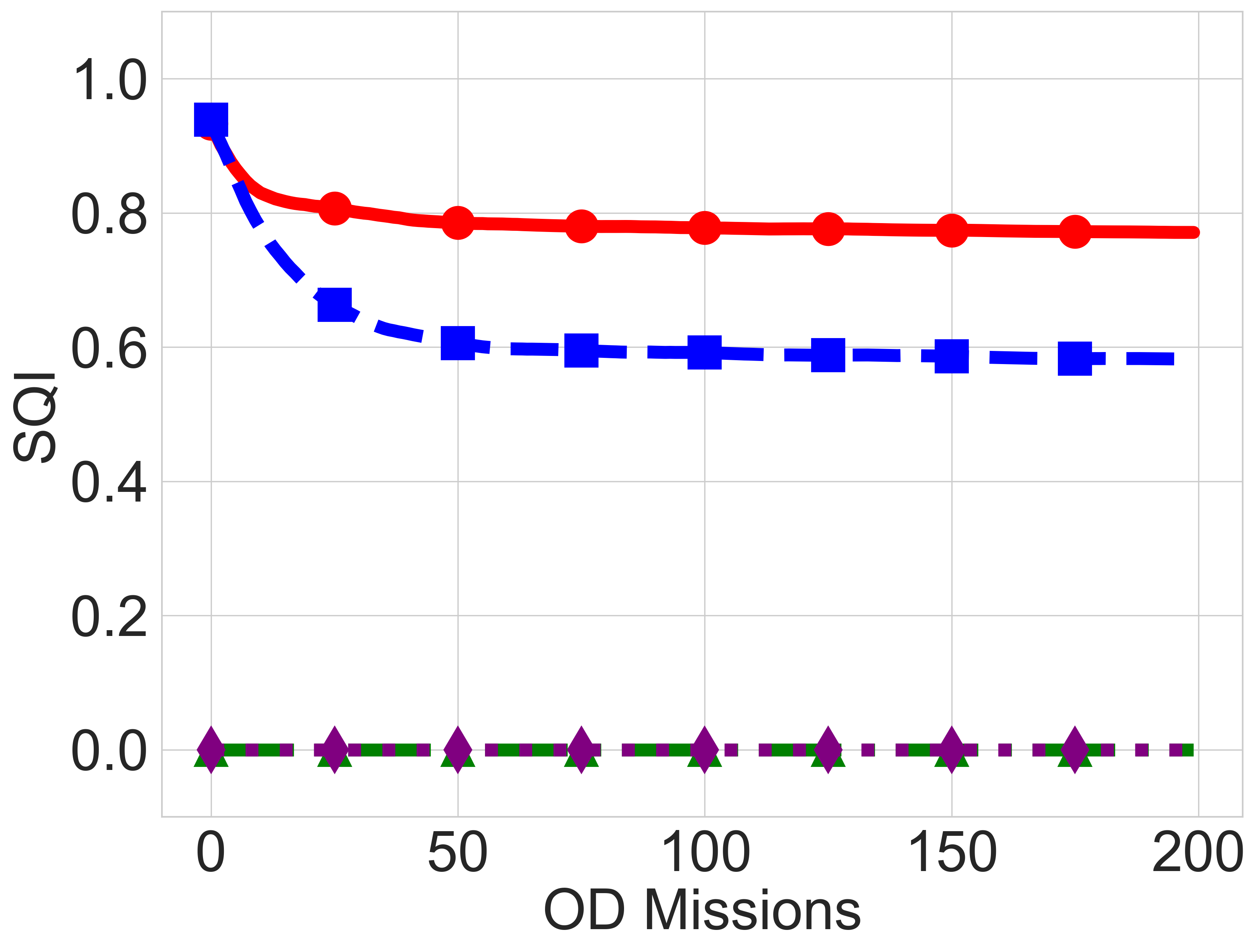}\label{fig:smm_quality_index}}
\hfil
\subfloat[SMM Coverage Index ($\text{SCI}$)]{\includegraphics[width=0.25\textwidth]{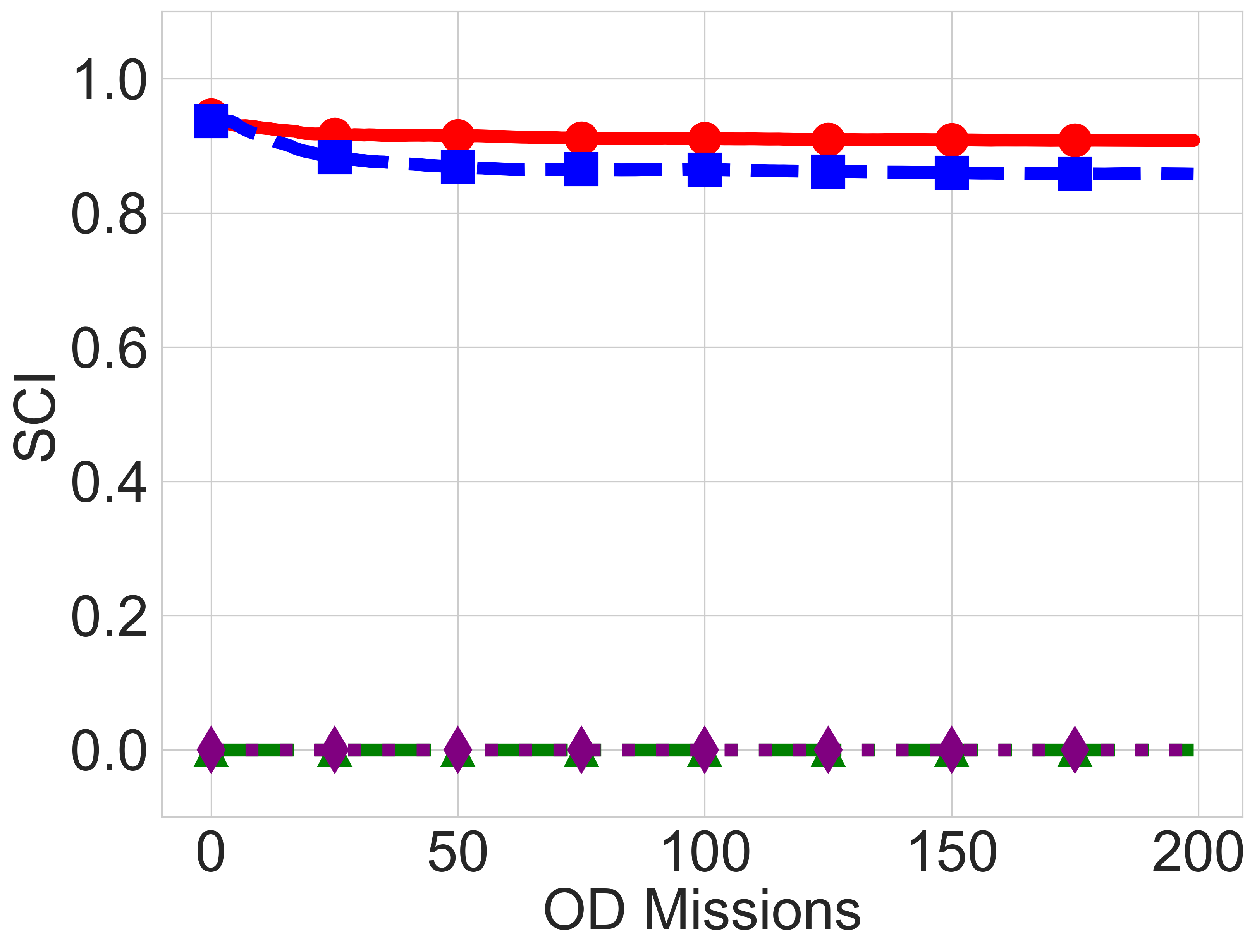}\label{fig:smm_coverage_index}}
\hfil
\caption{Performance comparison of four schemes under a fixed attack rate of 0.4.}
\label{fig:performance_analysis_0.4_attack_rate}
\vspace{-3mm}
\end{figure*}

\begin{figure*}[t]
\centering
\subfloat{\includegraphics[width=0.5\textwidth]{figures/metrics_legend.png}}
\hfil

\vspace{-3mm}
\setcounter{subfigure}{0}
\subfloat[UGV Compromised Rate]{\includegraphics[width=0.25\textwidth]{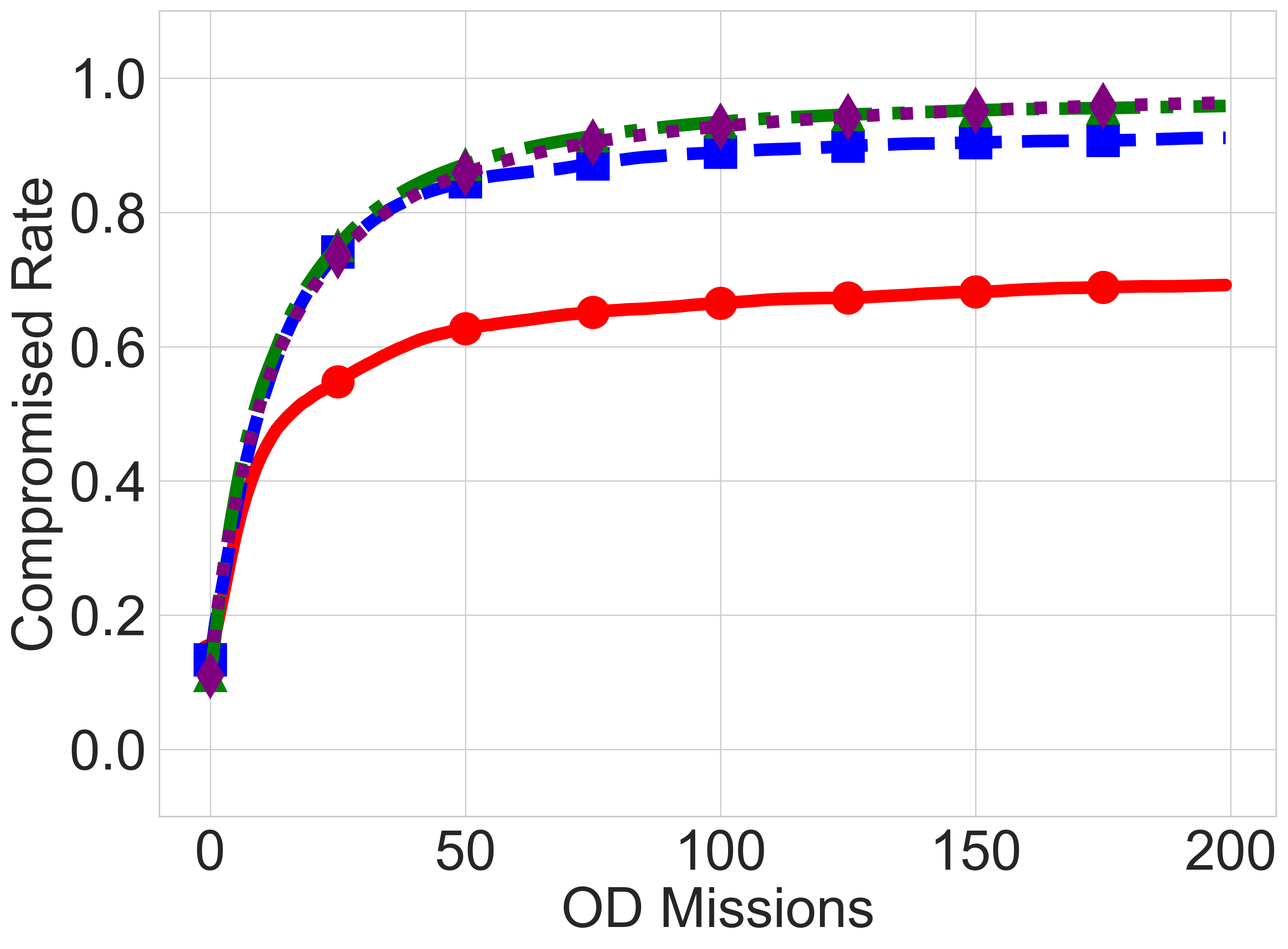}}
\hfil
\subfloat[AI Compromised Rate]{\includegraphics[width=0.25\textwidth]{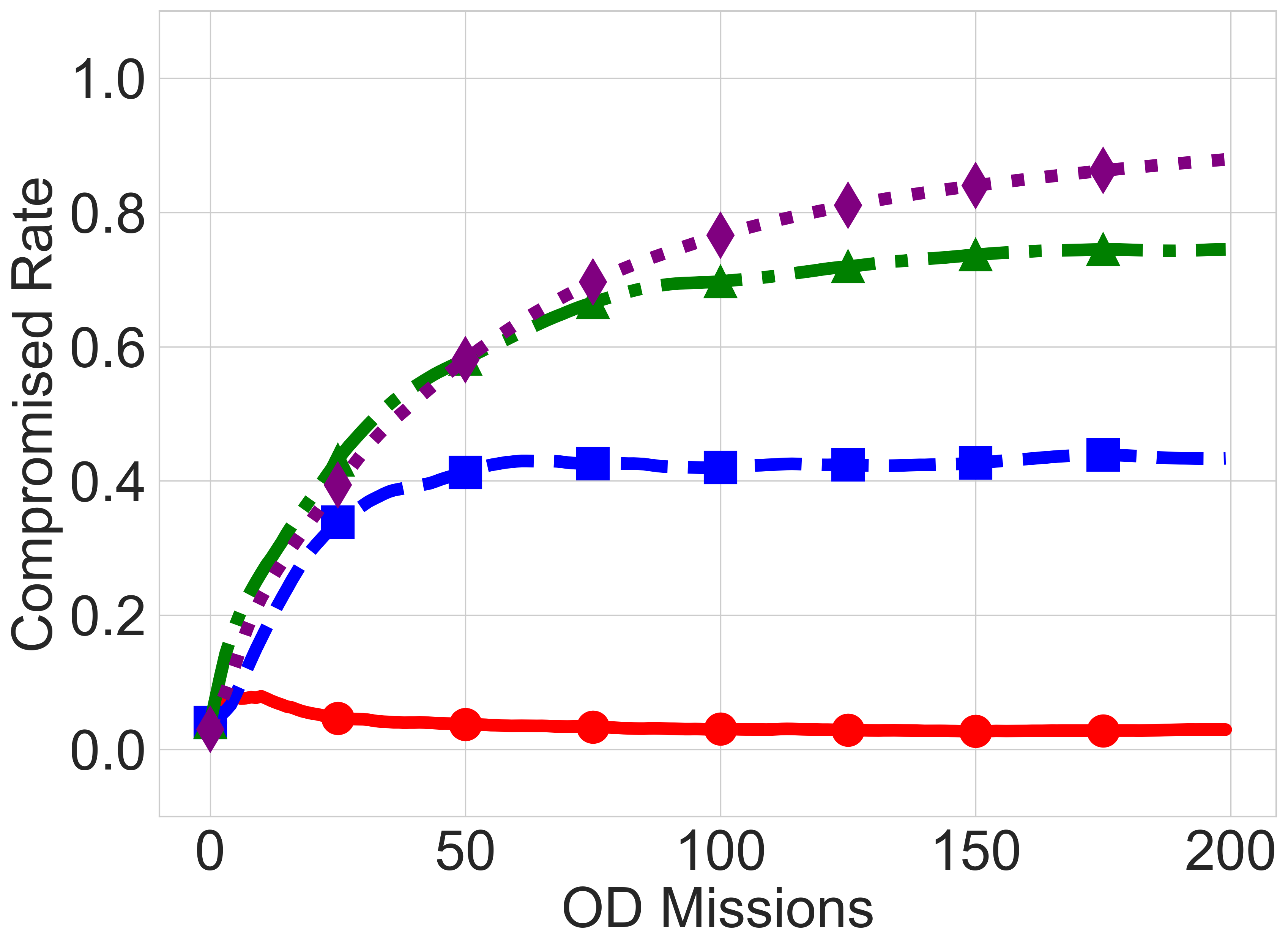}}
\hfil
\subfloat[Human Analysis Compromised Rate]{\includegraphics[width=0.25\textwidth]{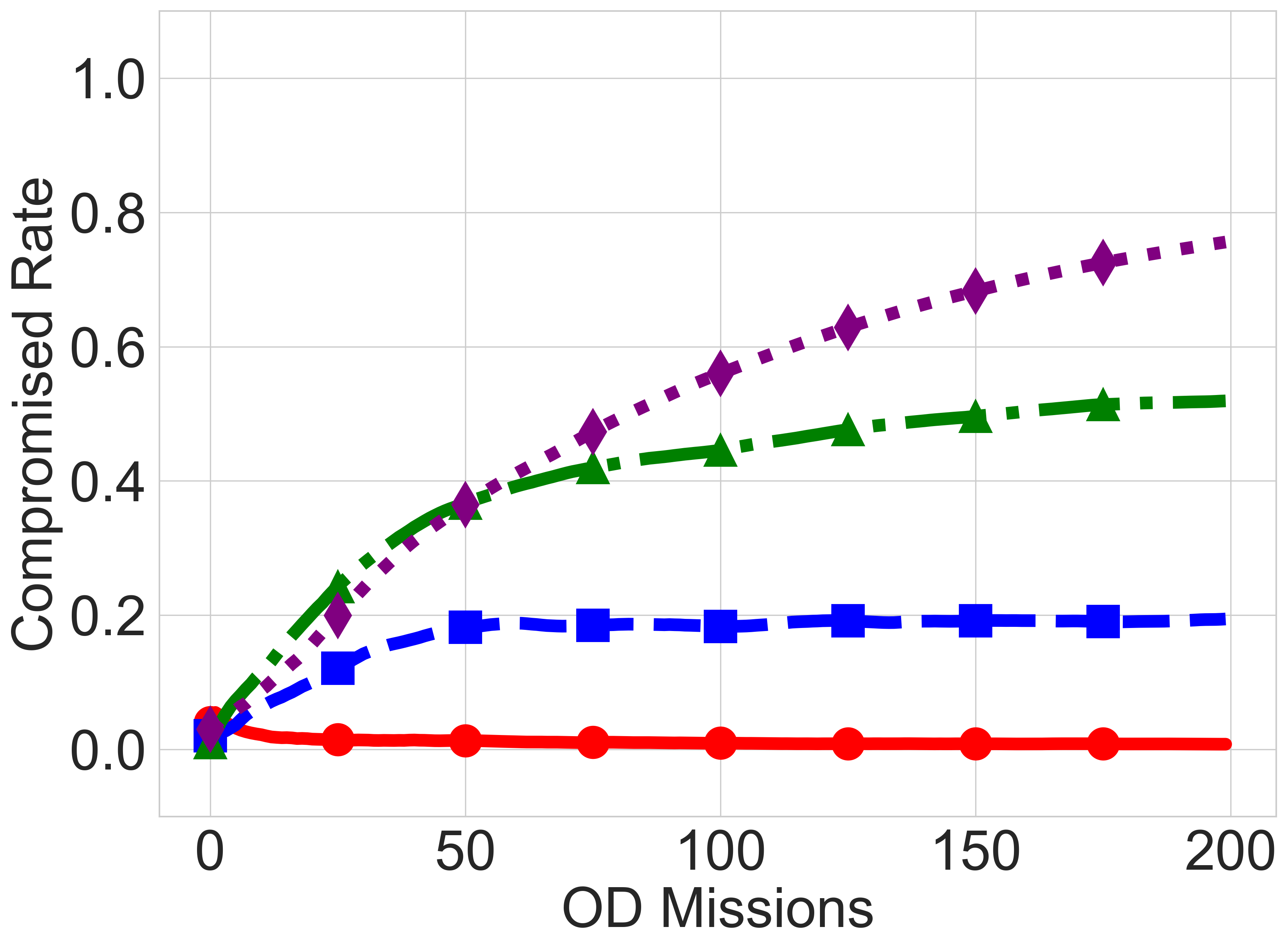}}
\hfil
\caption{Comparison of compromise rates for four schemes under a fixed attack rate of 0.4.}
\label{fig:compromised_members}
\vspace{-3mm}
\end{figure*}

\textbf{Fig.~\ref{fig:compromised_members}} shows compromise rates by role. DASH-DF had the lowest across UGVs, AI agents, and analysts, aligning with the MSR trend in \textbf{Fig.~\ref{fig:msr_missions}}. Rates rose early as attacks escalated, then stabilized as defenses responded. UGVs were most at risk due to exposed wireless links, AI agents faced a moderate threat, and analysts were least vulnerable due to resistance to social engineering.

\subsection{Sensitivity Analyses Under Varying Attack Rates} \label{sec:sensitivity_analysis_vary_attack_rate}

We conducted a sensitivity analysis by varying the attack rate from 0 to 1 in 0.2 increments, while fixing vulnerability parameters at $\mathrm{UGV:AI:Human = 0.3:0.1:0.05}$. Results at mission 200 confirm DASH’s robustness under escalating attack pressure and highlight the interaction between system components, defense mechanisms, and threat intensity.

\subsubsection{Mission Success Rate (MSR)} \label{sec:sensitivity_msr}
\textbf{Fig.~\ref{fig:sensitivity_msr_missions}} shows MSR trends across attack rates. DASH-DF consistently outperforms other schemes, maintaining $\sim$70\% success at the highest attack rate, while BASE drops below 10\%. The growing performance gap between DASH-DF and SMM-DF underscores the critical role of deception-based defenses. DF-only and BASE degrade sharply, especially at low attack rates (0.0--0.2), exposing weaknesses in non-adaptive defenses.

\subsubsection{SMM Quality Index (SQI)} \label{sec:sensitivity_sqi}
\textbf{Fig.~\ref{fig:sensitivity_smm_quality_index}} shows DASH-DF retains $\sim$70\% SQI at attack rate 1.0, while SMM-DF drops to $\sim$50\%. DASH-DF’s stable SQI reflects its resilience. DF-only and BASE remain at zero across all rates due to lacking effective information-sharing mechanisms.

\subsubsection{SMM Coverage Index (SCI)} \label{sec:sensitivity_sci}
\begin{figure*}[t]
\centering
\subfloat{\includegraphics[width=0.5\textwidth]{figures/metrics_legend.png}}
\hfil

\vspace{-3mm}
\setcounter{subfigure}{0}
\subfloat[Mission Success Rate ($\text{MSR}$)]{\includegraphics[width=0.25\textwidth]{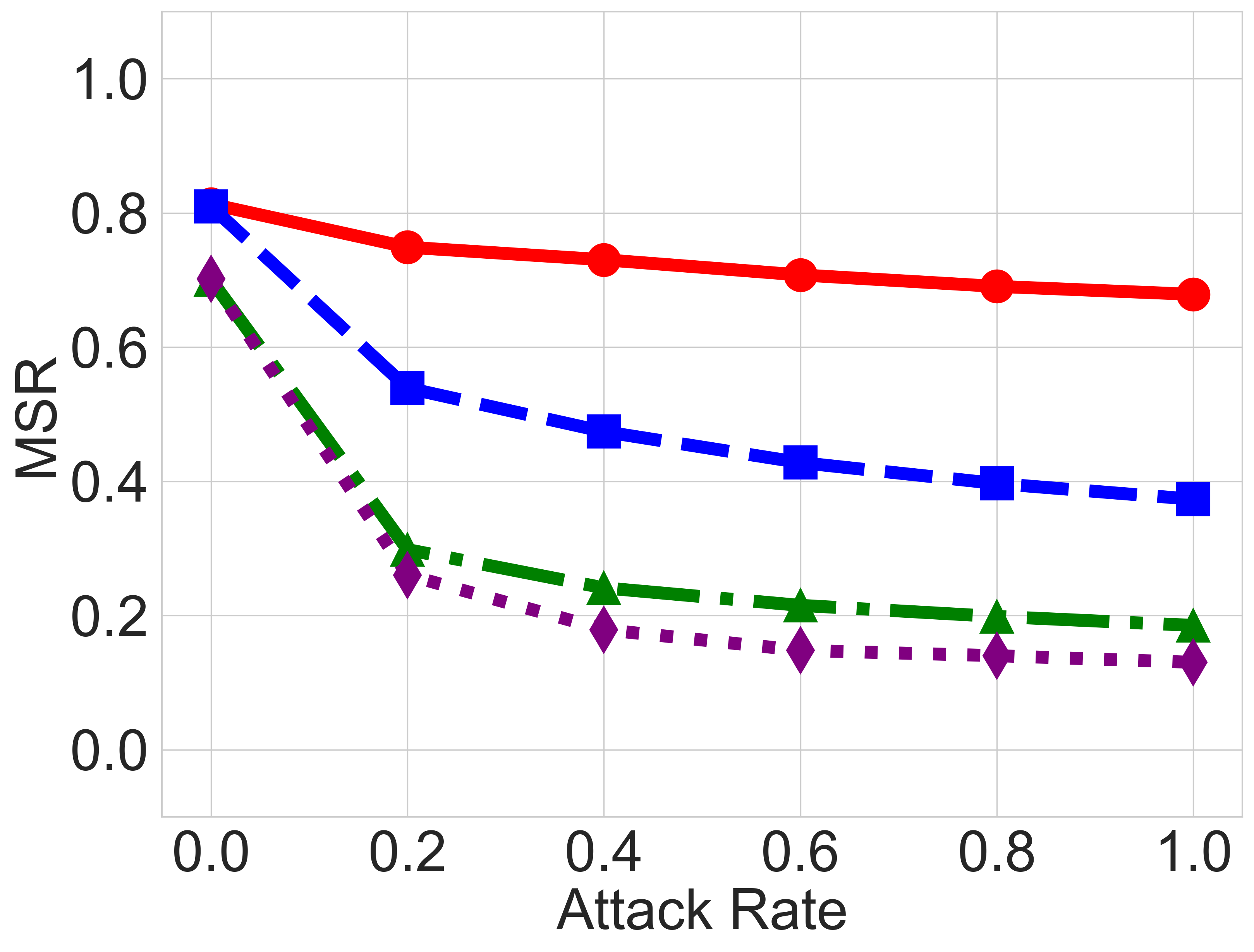}\label{fig:sensitivity_msr_missions}}
\hfil
\subfloat[SMM Quality Index ($\text{SQI}$)]{\includegraphics[width=0.25\textwidth]{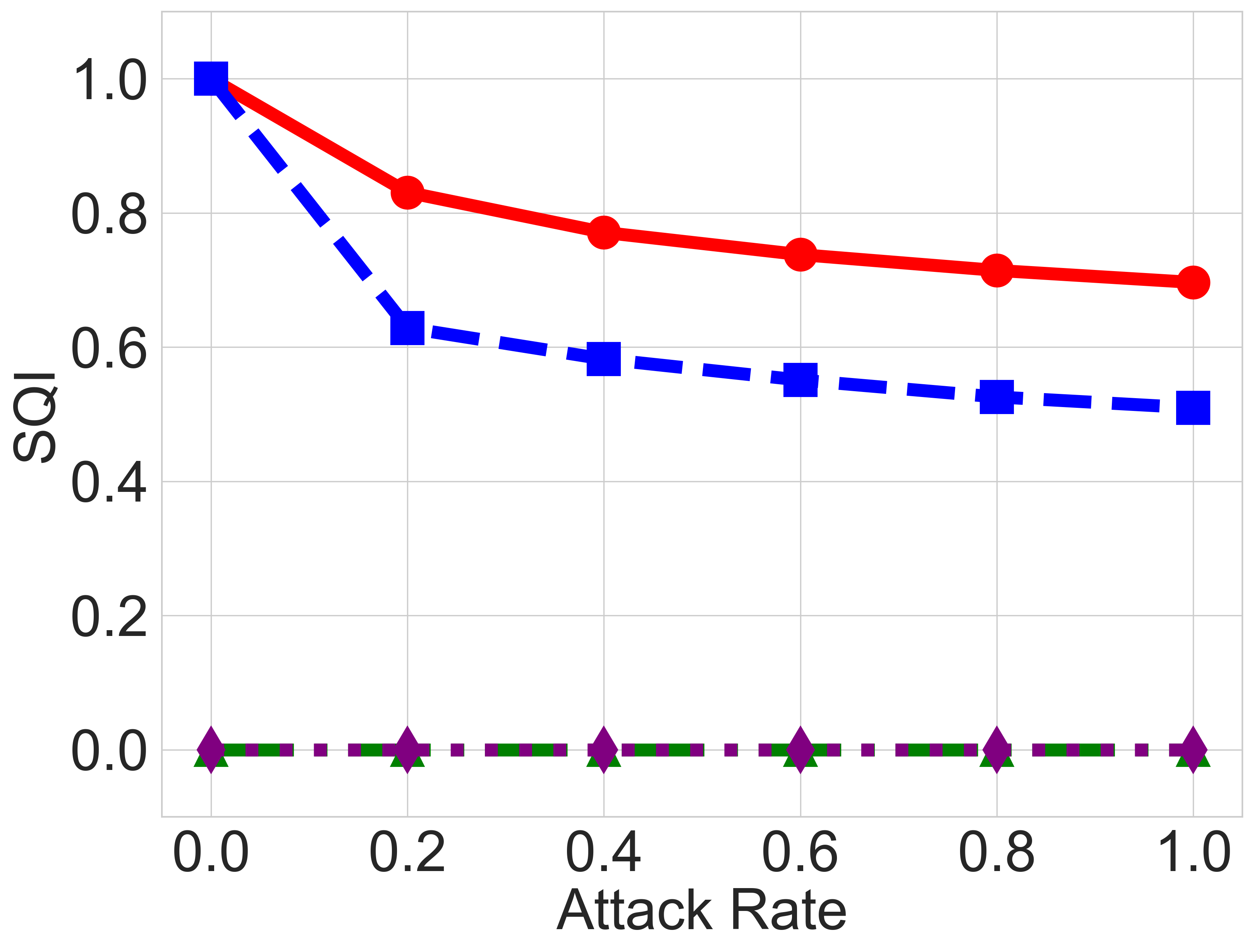}\label{fig:sensitivity_smm_quality_index}}
\hfil
\subfloat[SMM Coverage Index ($\text{SCI}$)]{\includegraphics[width=0.25\textwidth]{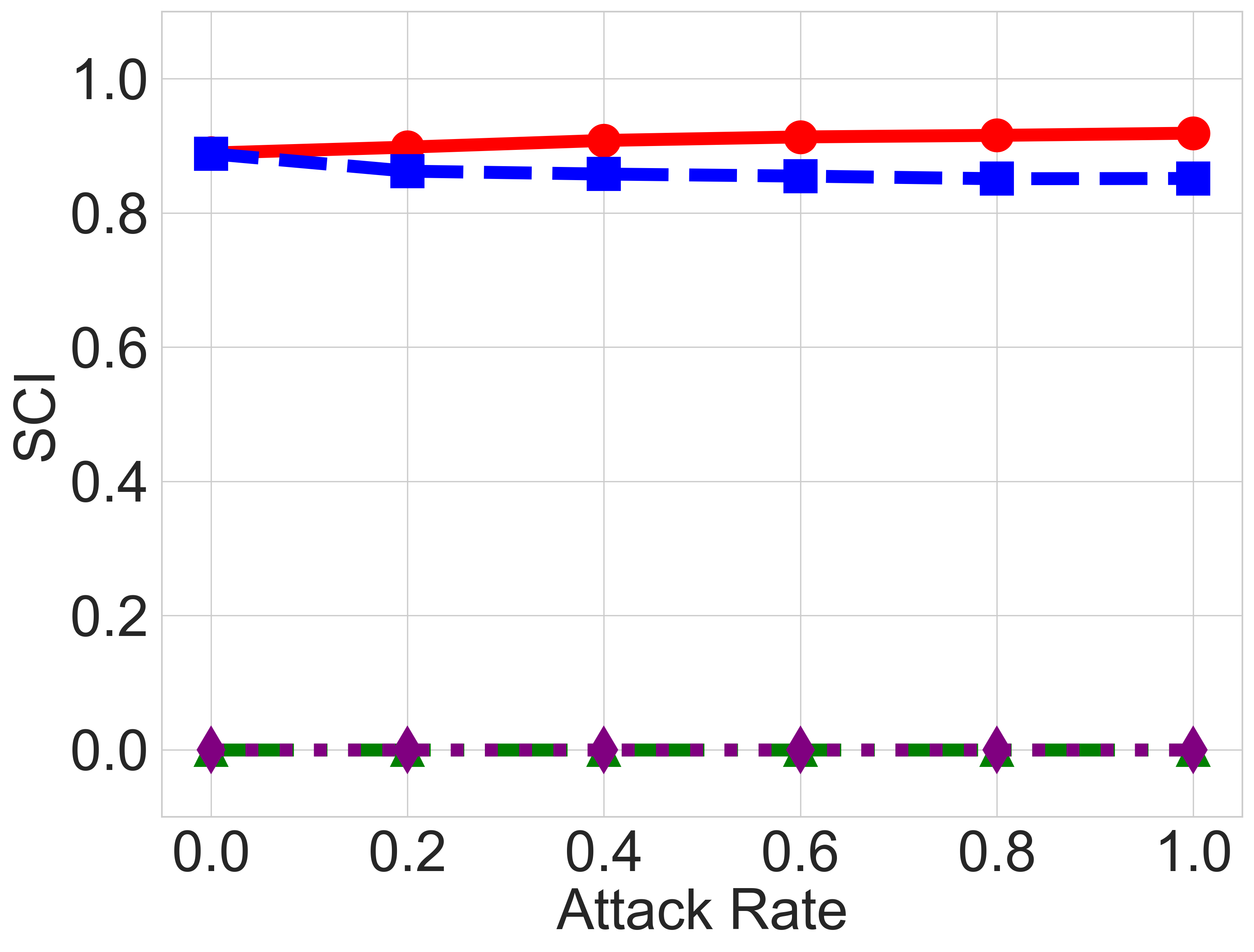}\label{fig:sensitivity_smm_coverage_index}}
\hfil
\caption{Performance analysis of four schemes across varying attack rates (0.0 to 1.0).}
\label{fig:sensitivity_analysis}
\vspace{-3mm}
\end{figure*}

\textbf{Fig.~\ref{fig:sensitivity_smm_coverage_index}} shows that DASH-DF sustains SCI near 0.9 across all attack rates. SMM-DF declines slightly to $\sim$0.85, highlighting ADTM’s role in maintaining communication. DF-only and BASE remain at zero, lacking a collaborative sharing infrastructure.

\subsubsection{Operational Costs} \label{sec:sensitivity_operational_costs}
\begin{figure*}[t]
\centering
\subfloat{\includegraphics[width=0.5\textwidth]{figures/metrics_legend.png}}
\hfil

\vspace{-3mm}
\setcounter{subfigure}{0}
\subfloat[UGV Operational Cost]{\includegraphics[width=0.25\textwidth]{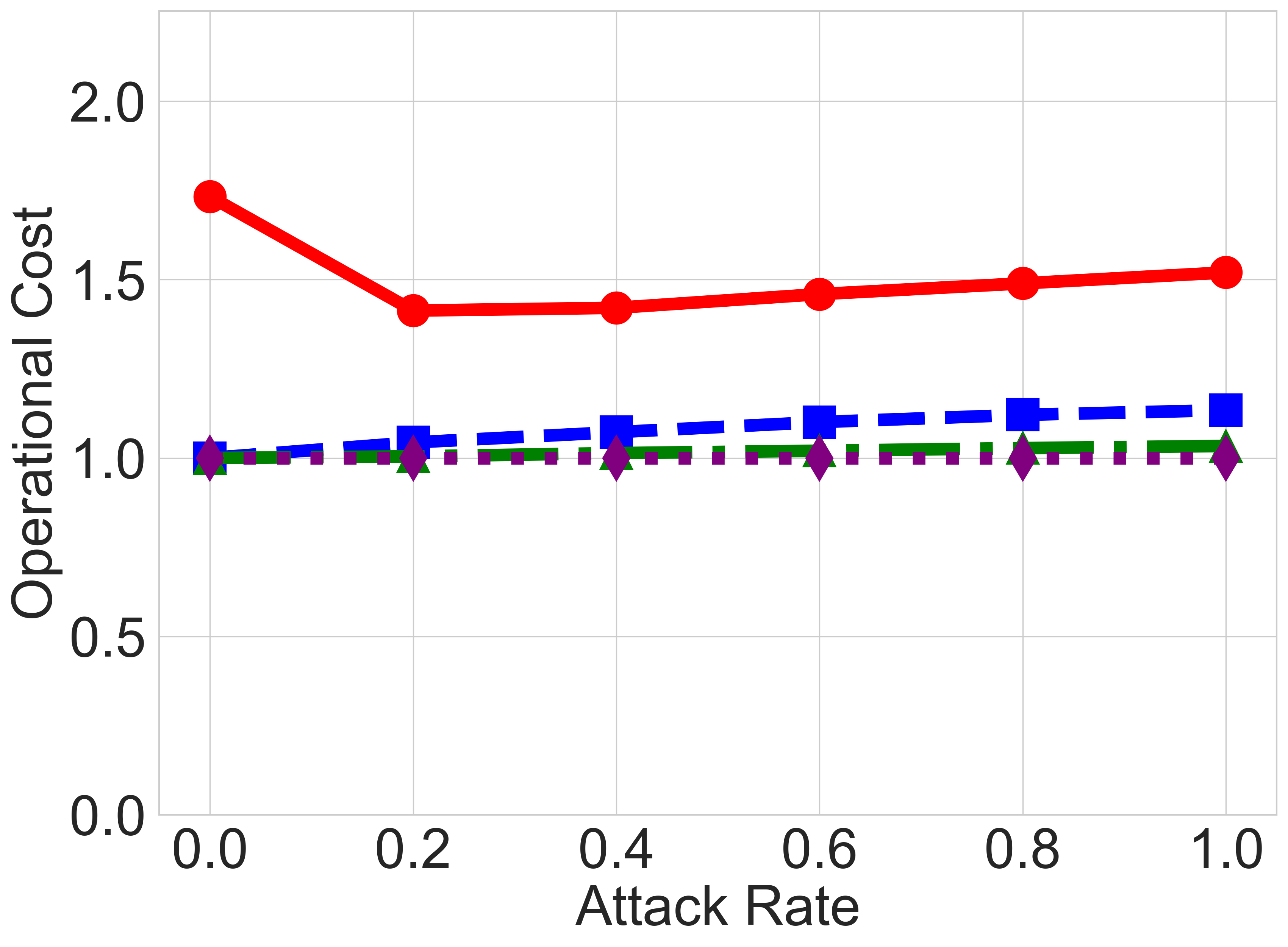}}
\hfil
\subfloat[AI Operational Cost]{\includegraphics[width=0.25\textwidth]{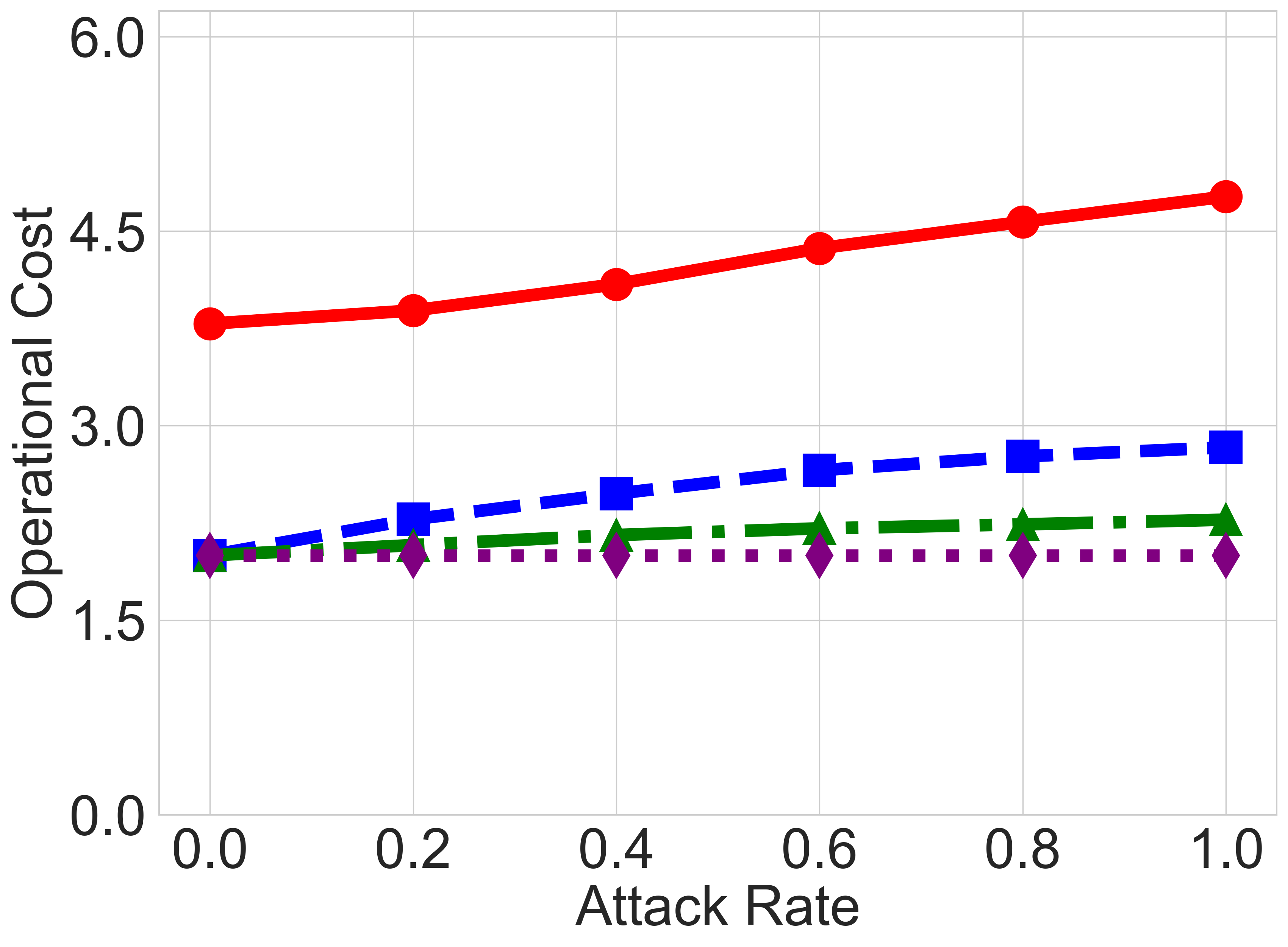}}
\hfil
\subfloat[Human Analysis Operational Cost]{\includegraphics[width=0.25\textwidth]{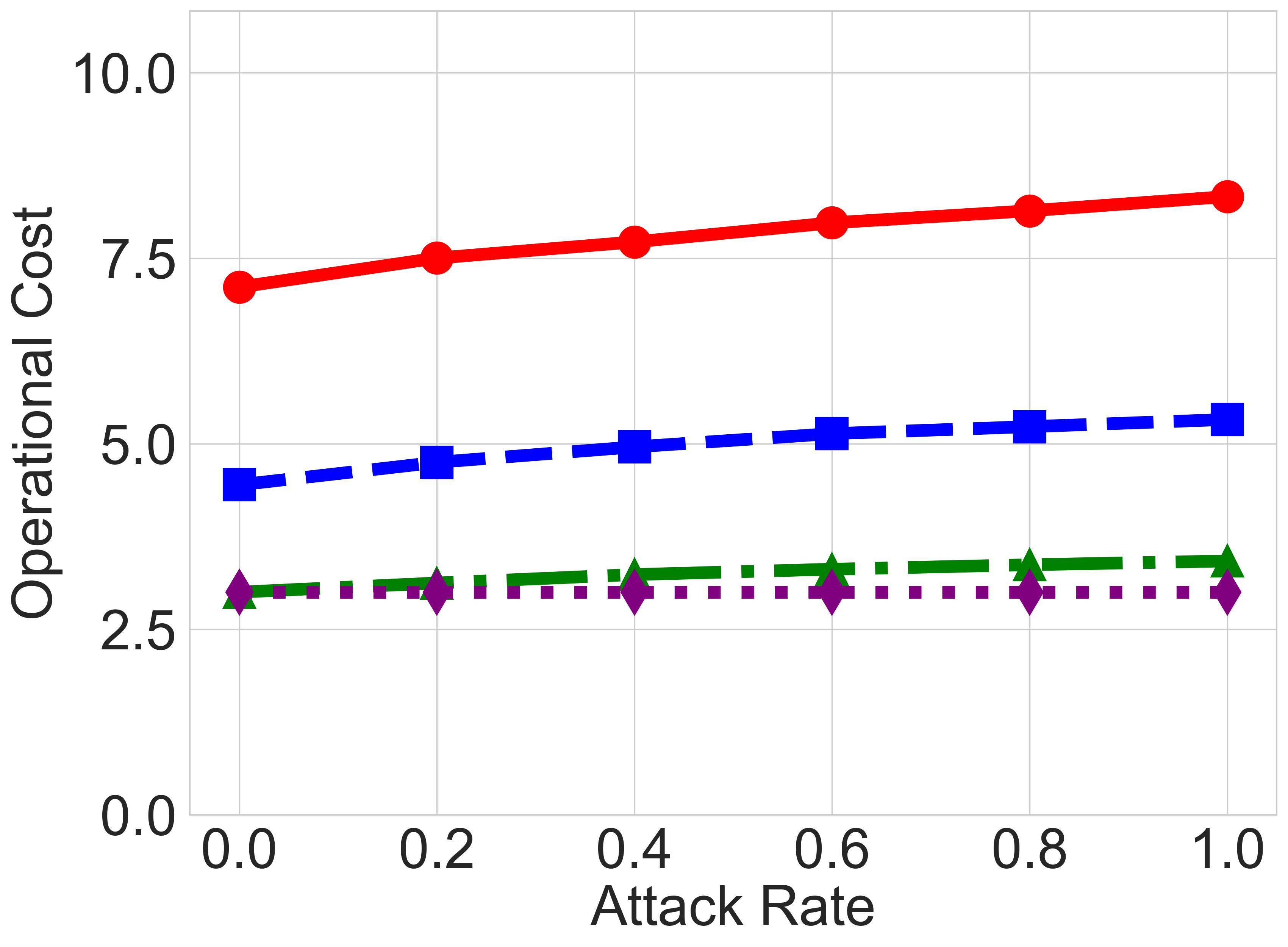}}
\hfil
\subfloat[Total Operational Cost of All Members]{\includegraphics[width=0.25\textwidth]{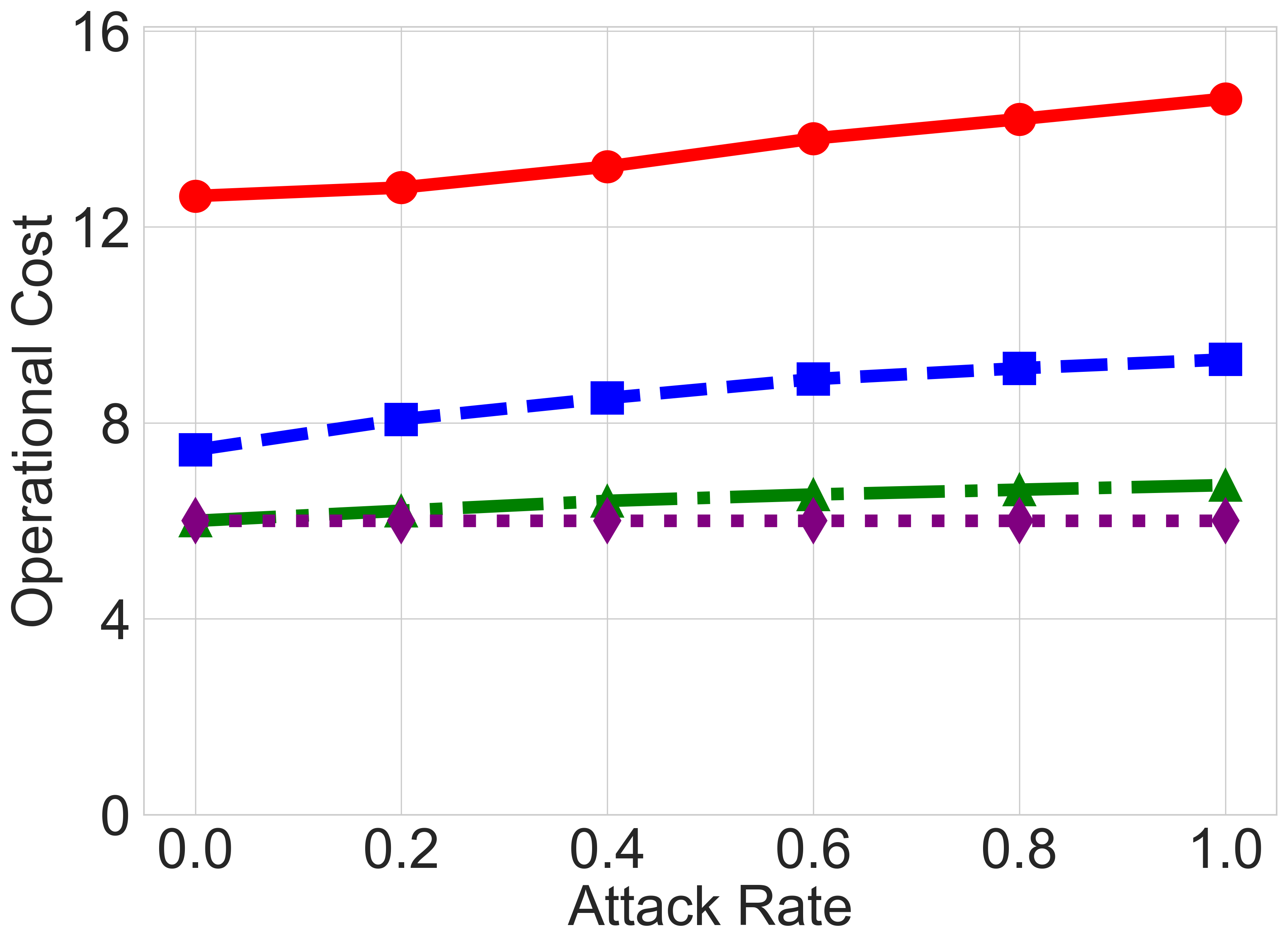}}
\hfil
\caption{Operational cost analysis of four schemes across varying attack rates (0.0 to 1.0).}
\label{fig:sensitivity_operational_cost}
\vspace{-3mm}
\end{figure*}

\textbf{Fig.~\ref{fig:sensitivity_operational_cost}} presents cost trends. DASH-DF incurs higher UGV costs (1.4–1.7) due to bait tasks and reinstallations. Although other schemes show lower AI and human costs, they suffer higher compromise rates (\textbf{Fig.~\ref{fig:sensitivity_compromised_members}}). This tradeoff underscores DASH-DF’s investment in resilience.

\subsubsection{Ratio of Compromised Members} \label{sec:sensitivity_compromised_members}
\begin{figure*}[t]
\centering
\subfloat{\includegraphics[width=0.5\textwidth]{figures/metrics_legend.png}}
\hfil

\vspace{-3mm}
\setcounter{subfigure}{0}
\subfloat[UGV Compromised Rate]{\includegraphics[width=0.25\textwidth]{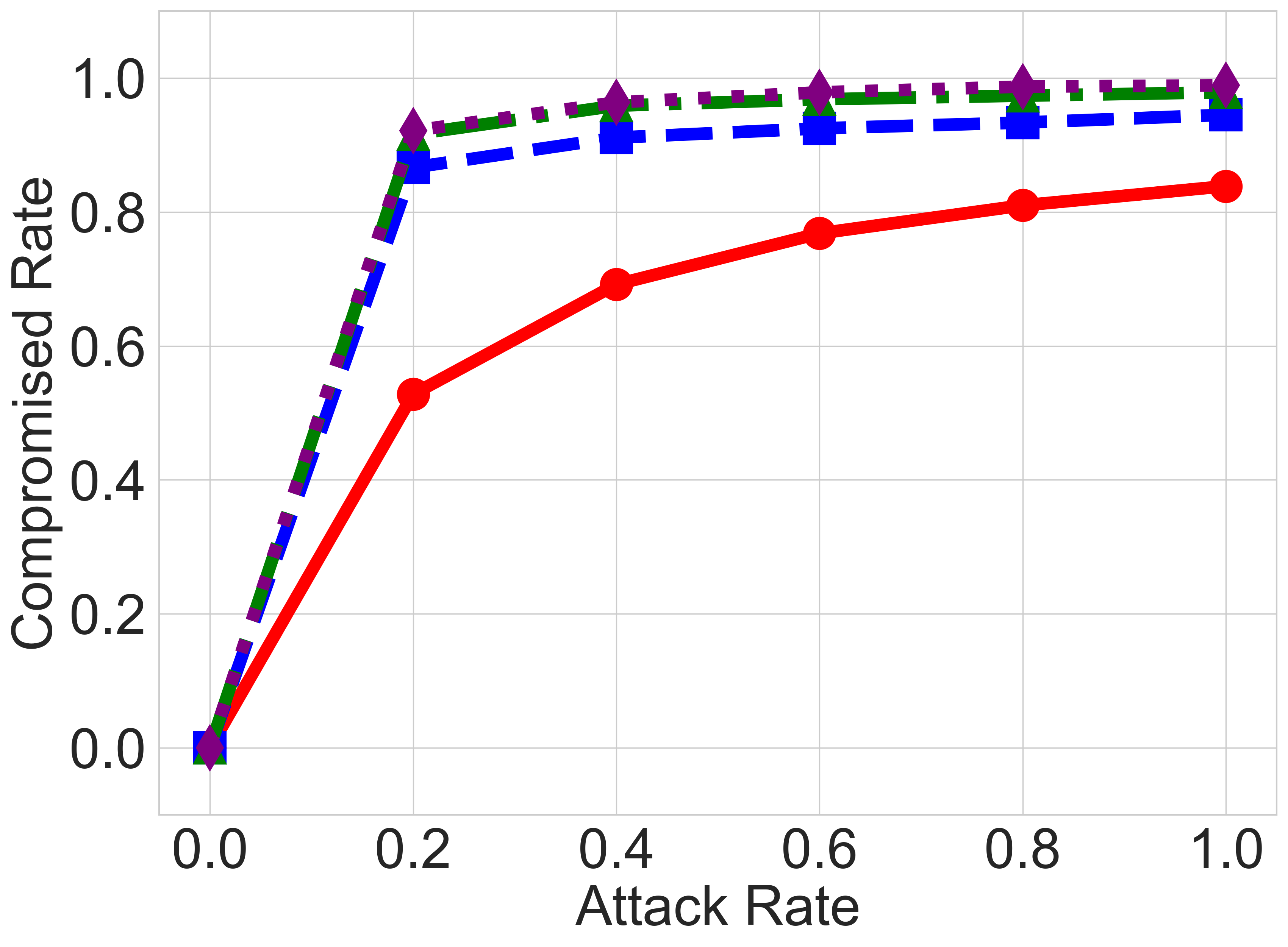}}
\hfil
\subfloat[AI Compromised Rate]{\includegraphics[width=0.25\textwidth]{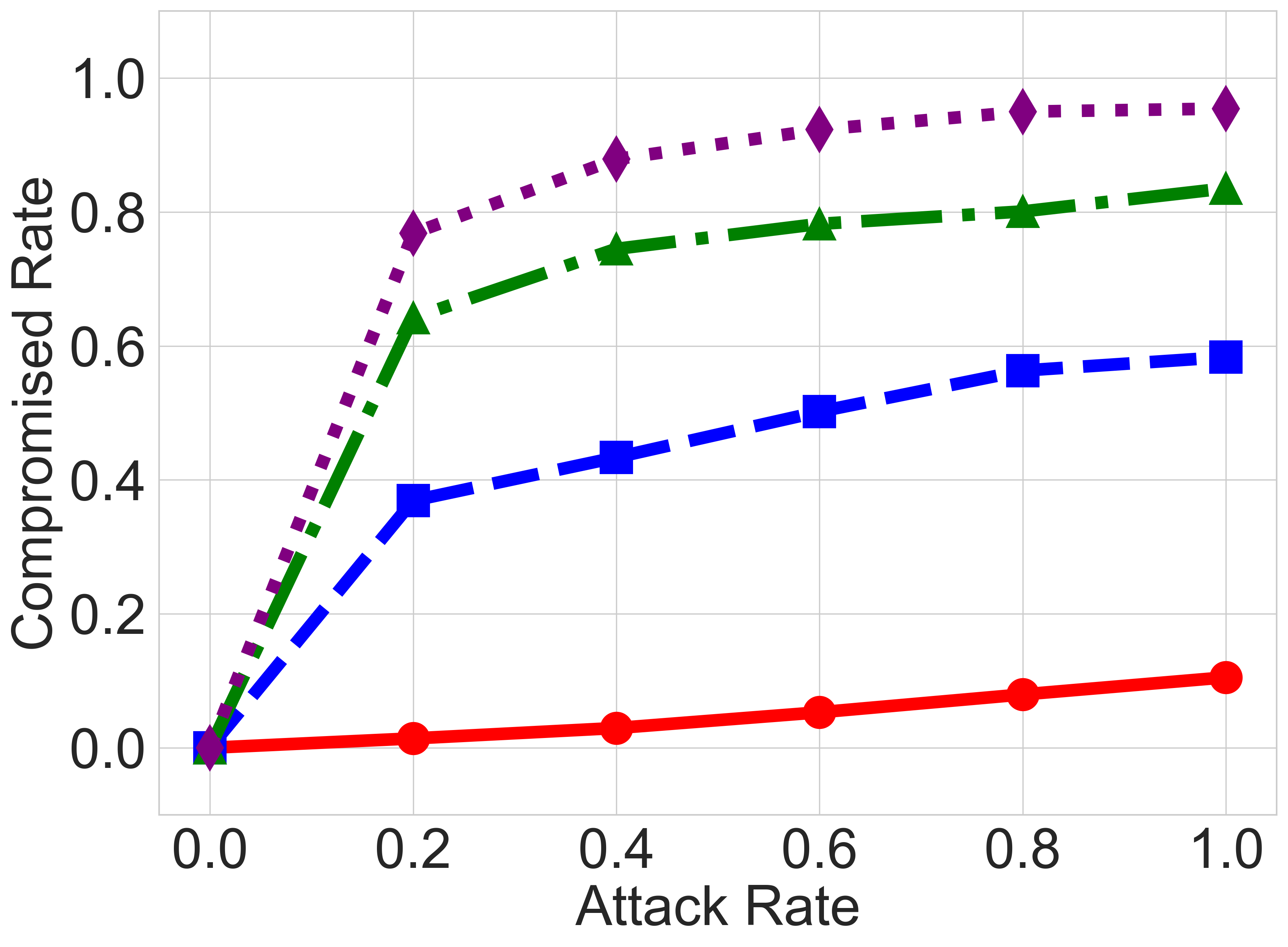}}
\hfil
\subfloat[Human Analysis Compromised Rate]{\includegraphics[width=0.25\textwidth]{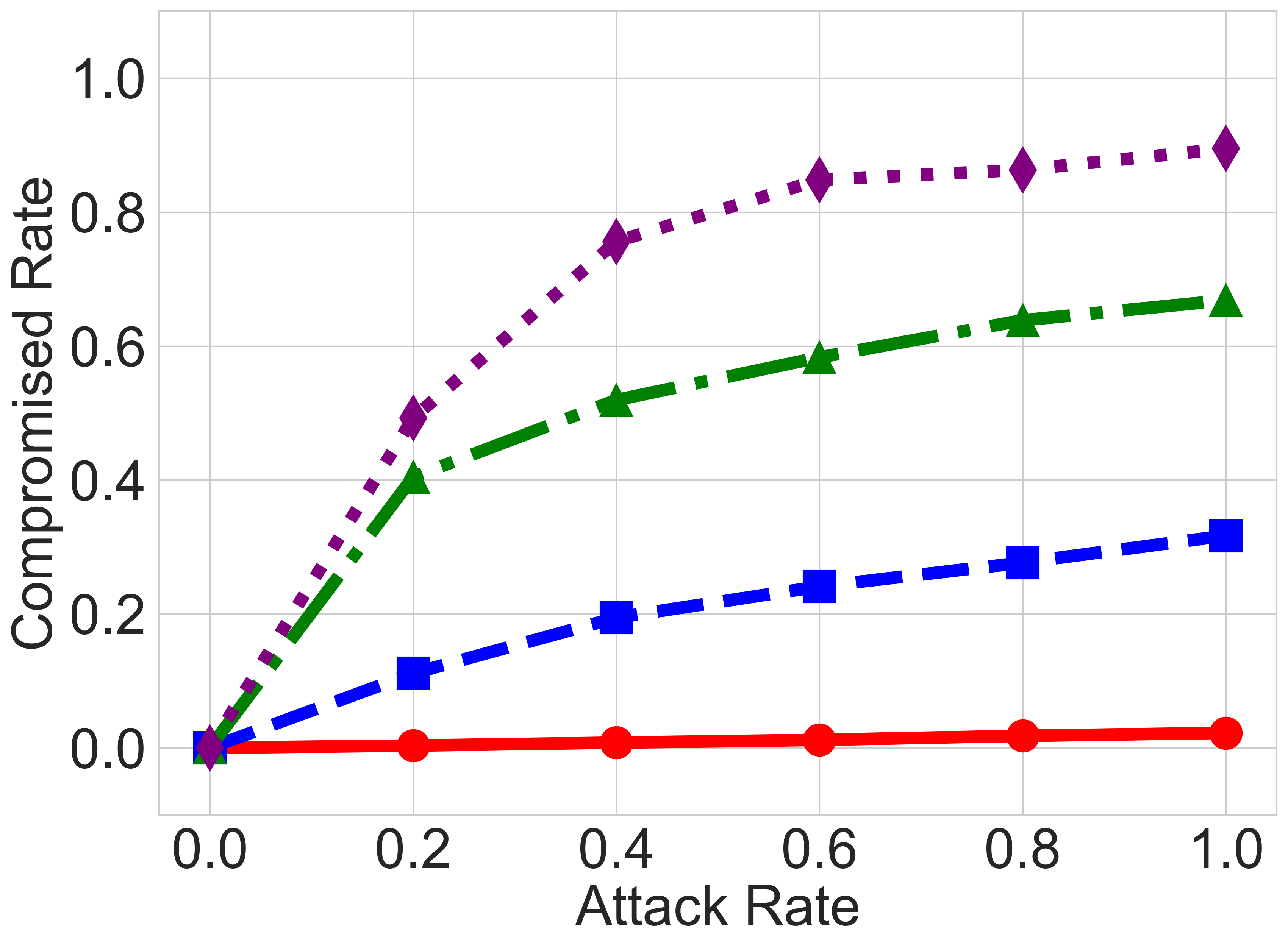}}
\hfil
\caption{Compromise rate analysis of four schemes across varying attack rates (0.0 to 1.0).}
\label{fig:sensitivity_compromised_members}
\vspace{-3mm}
\end{figure*}

\textbf{Fig.~\ref{fig:sensitivity_compromised_members}} shows that SMM-DF, DF-only, and BASE experience sharp compromise increases at low attack rates (0–0.2). In contrast, DASH-DF maintains low compromise levels, even at an attack rate of 1.0, AI remains below 0.15, and human analysts under 0.1, demonstrating the effectiveness of deception-based protections.

\subsection{Sensitivity Analyses Under Varying Member Vulnerability}
Beyond attack frequency, we analyzed how varying team member vulnerabilities affect system performance. UGV vulnerability increased from 0.2 to 0.6 (in 0.1 increments), with AI vulnerability set to half the UGV value and human analyst vulnerability to half the AI value, yielding triplets \{UGV:AI:Human\} = \{0.2:0.05:0.02, 0.3:0.1:0.05, 0.4:0.2:0.1, 0.5:0.3:0.15, 0.6:0.4:0.2\}. Results under a fixed attack rate of 0.4 are shown in Figs.~\ref{fig:vulnerability_sensitivity_analysis}--\ref{fig:vulnerability_sensitivity_compromised_members}.

\subsubsection{Mission Success Rate (MSR)} \label{sec:vulnerability_sensitivity_msr}
\textbf{Fig.~\ref{fig:vulnerability_sensitivity_msr_missions}} shows that as vulnerabilities increase, MSR declines across all schemes. DASH-DF remains most resilient, sustaining over 80\% success even at peak vulnerability, highlighting the benefit of combining SMM and deception. SMM-DF drops to $\sim$60\%, while DF-only and BASE fall sharply to $\sim$20\% and $\sim$10\%, respectively, failing to address growing compromise risks.

\subsubsection{SMM Quality Index (SQI)} \label{sec:vulnerability_sensitivity_sqi}
\textbf{Fig.~\ref{fig:vulnerability_sensitivity_smm_quality_index}} shows DASH-DF maintains $\sim$0.75 SQI at peak vulnerability, preserving integrity. SMM-DF drops to $\sim$0.6, while DF-only and BASE stay at zero due to lacking information-sharing.

\subsubsection{SMM Coverage Index (SCI)} \label{sec:vulnerability_sensitivity_sci}
\begin{figure*}[t]
\centering
\subfloat{\includegraphics[width=0.5\textwidth]{figures/metrics_legend.png}}
\hfil

\vspace{-3mm}
\setcounter{subfigure}{0}
\subfloat[Mission Success Rate ($\text{MSR}$)]{\includegraphics[width=0.25\textwidth]{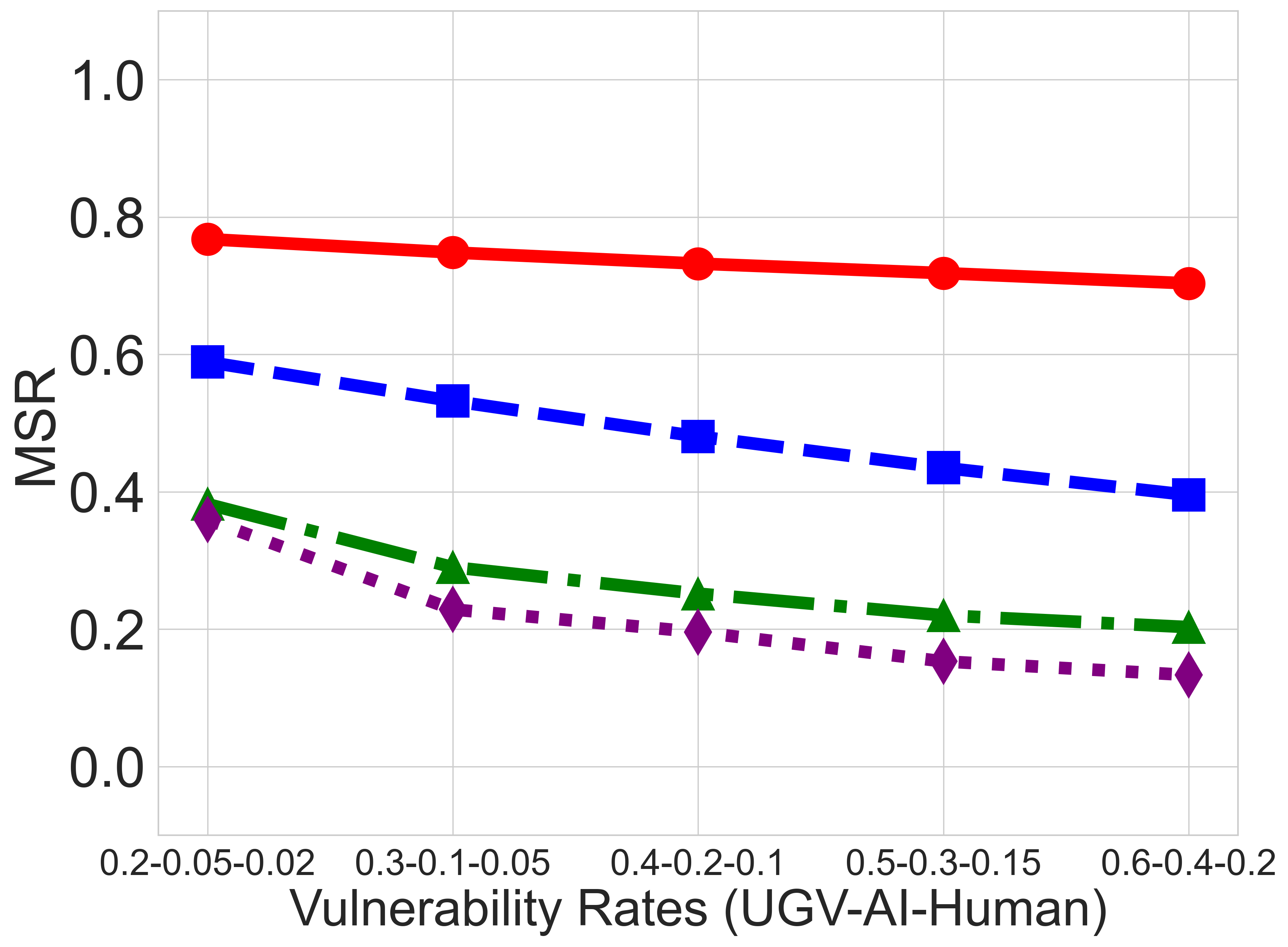}\label{fig:vulnerability_sensitivity_msr_missions}}
\hfil
\subfloat[SMM Quality Index ($\text{SQI}$)]{\includegraphics[width=0.25\textwidth]{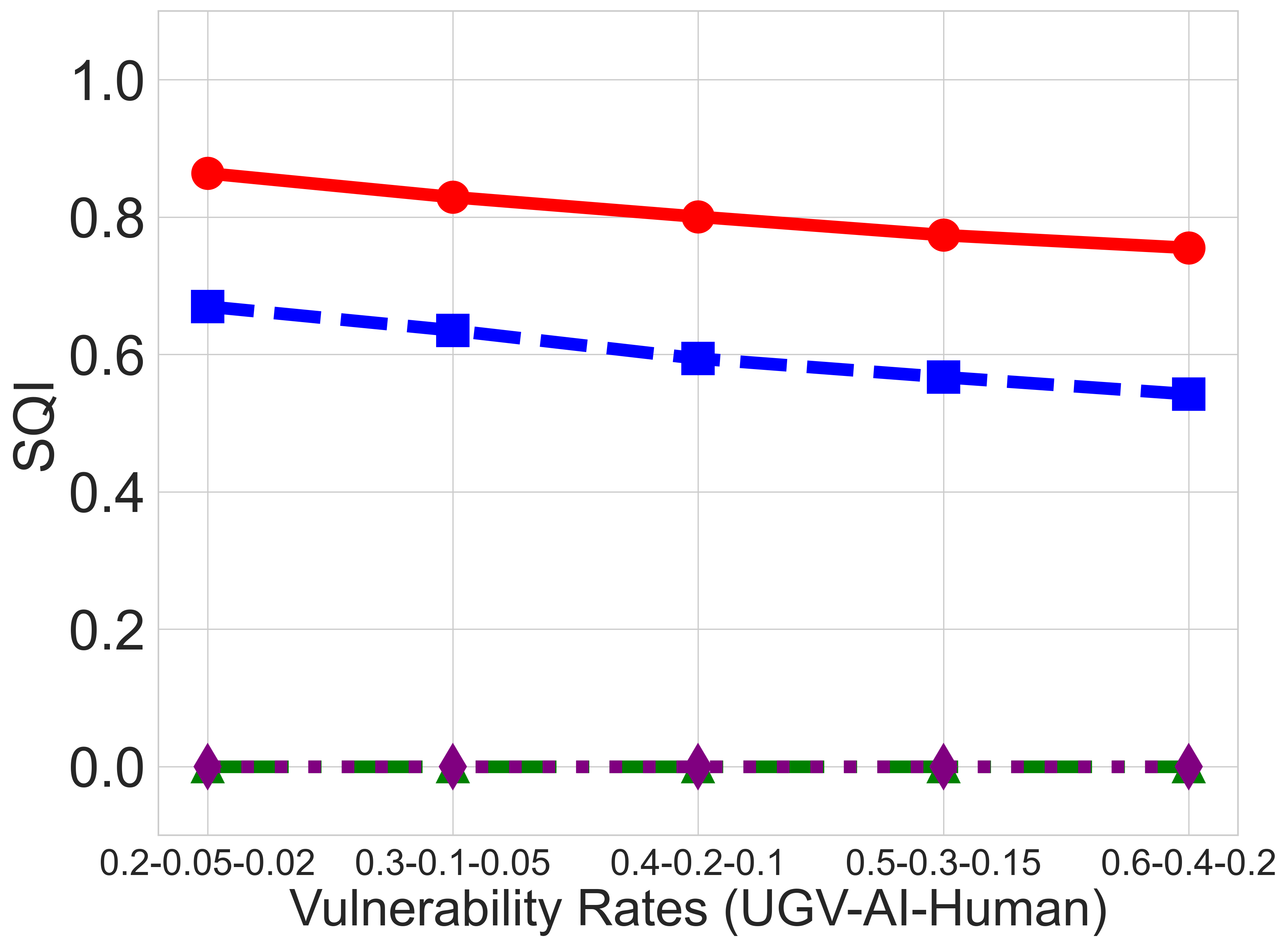}\label{fig:vulnerability_sensitivity_smm_quality_index}}
\hfil
\subfloat[SMM Coverage Index ($\text{SCI}$)]{\includegraphics[width=0.25\textwidth]{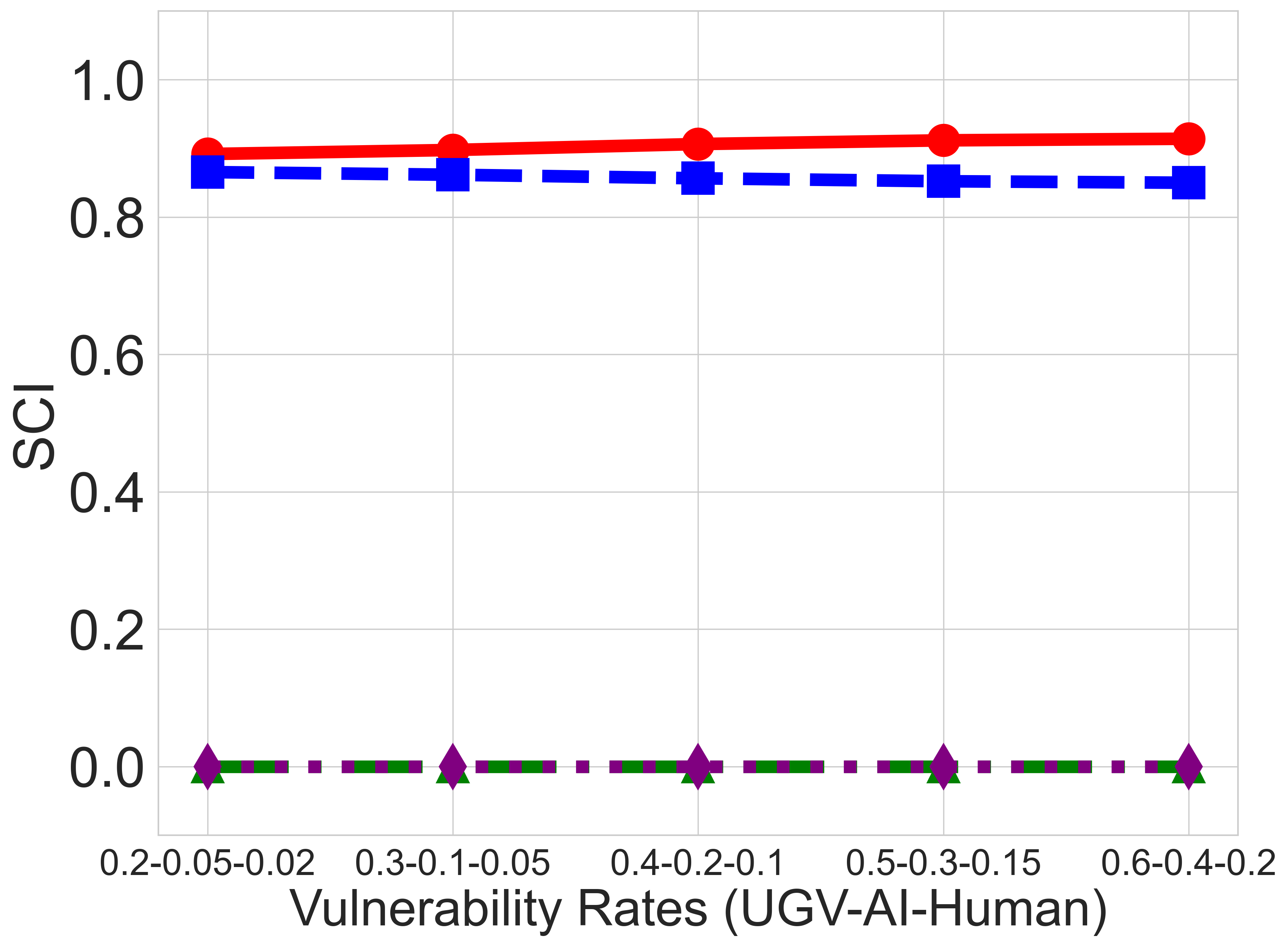}\label{fig:vulnerability_sensitivity_smm_coverage_index}}
\hfil
\caption{Performance analysis of four schemes under varying attack rates (0.0 to 1.0).}
\label{fig:vulnerability_sensitivity_analysis}
\vspace{-3mm}
\end{figure*}

As shown in \textbf{Fig.~\ref{fig:vulnerability_sensitivity_smm_coverage_index}}, DASH-DF maintains $\sim$0.9 SCI across all settings. SMM-DF declines from $\sim$0.95 to $\sim$0.70, reflecting reduced information flow with higher compromise. The widening gap underscores deception’s role in preserving communication completeness.

\subsubsection{Operational Costs} \label{sec:vulnerability_sensitivity_operational_costs}
\begin{figure*}[t]
\centering
\subfloat{\includegraphics[width=0.5\textwidth]{figures/metrics_legend.png}}
\hfil

\vspace{-3mm}
\setcounter{subfigure}{0}
\subfloat[UGV Operational Cost]{\includegraphics[width=0.25\textwidth]{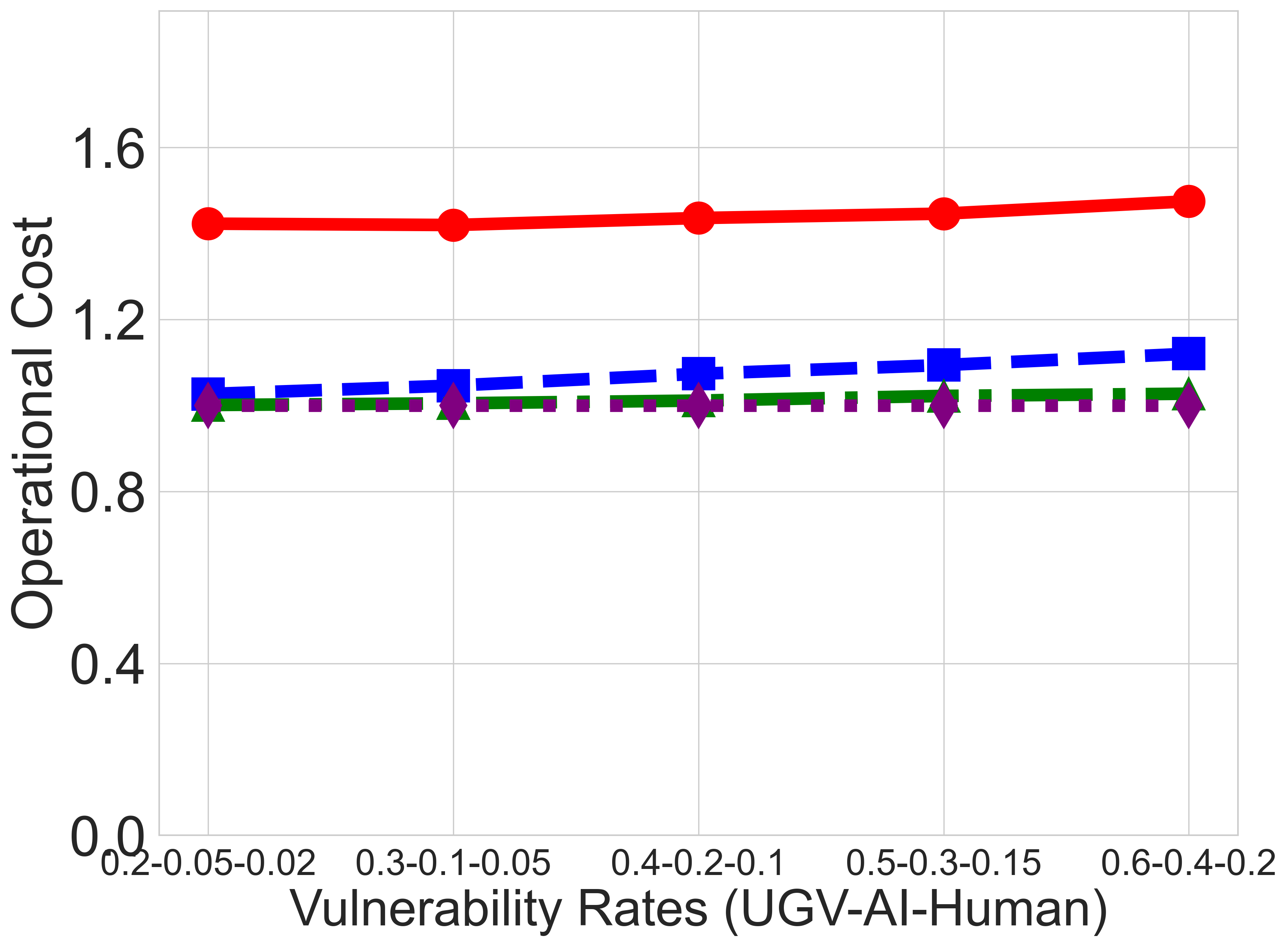}}
\hfil
\subfloat[AI Operational Cost]{\includegraphics[width=0.25\textwidth]{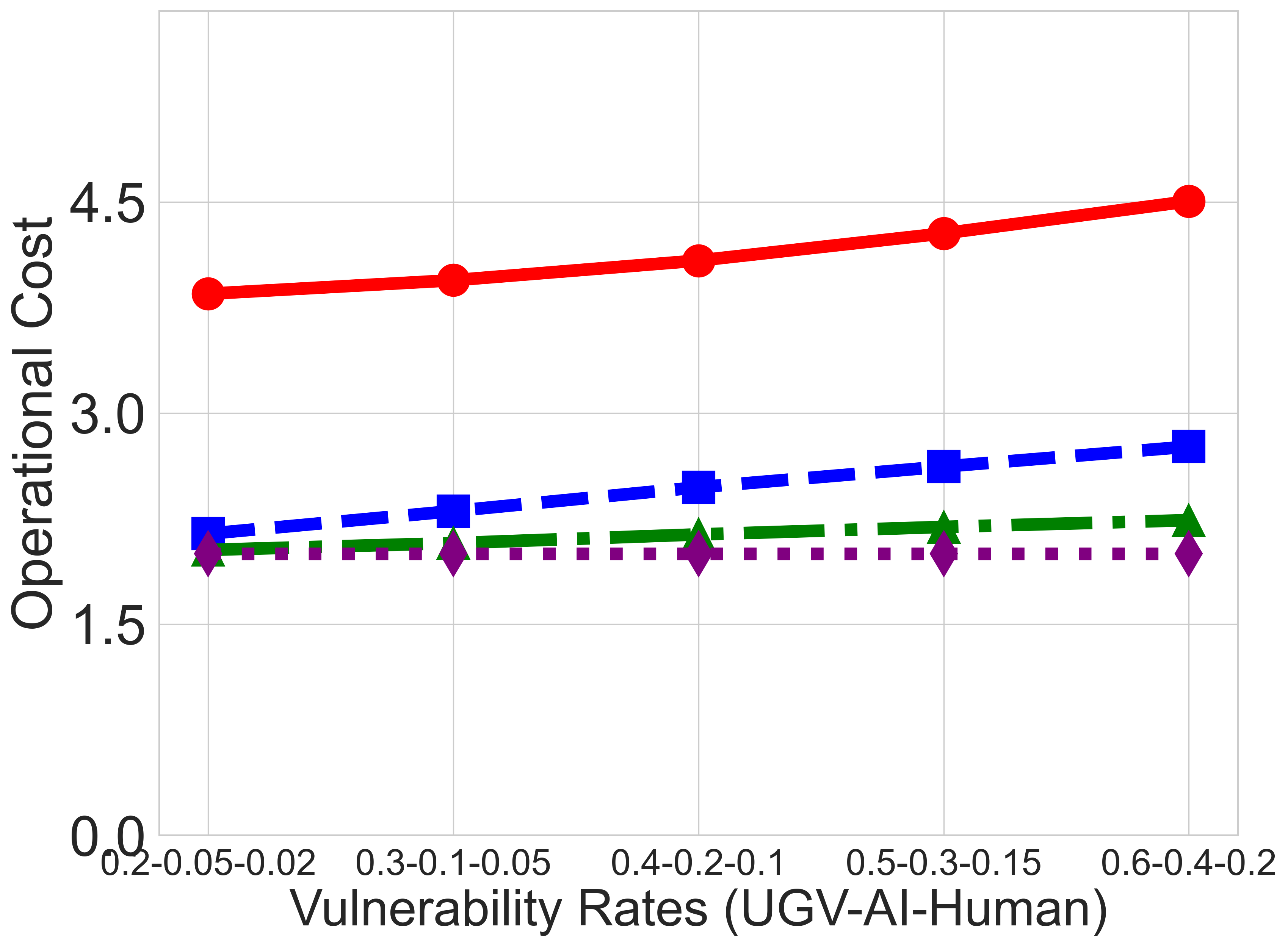}}
\hfil
\subfloat[Human Analysis Operational Cost]{\includegraphics[width=0.25\textwidth]{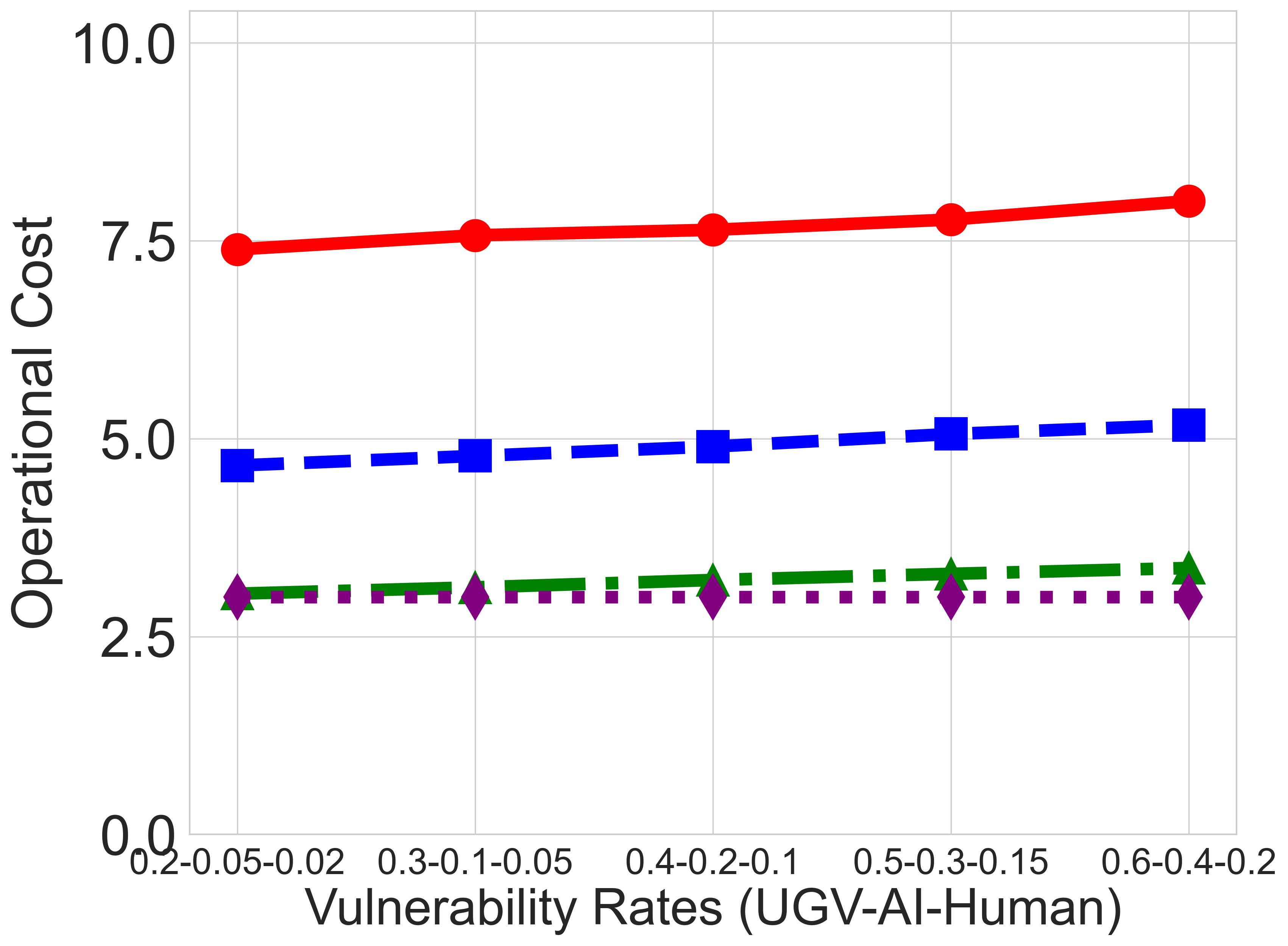}}
\hfil
\subfloat[Total Operational Cost of All Members]{\includegraphics[width=0.25\textwidth]{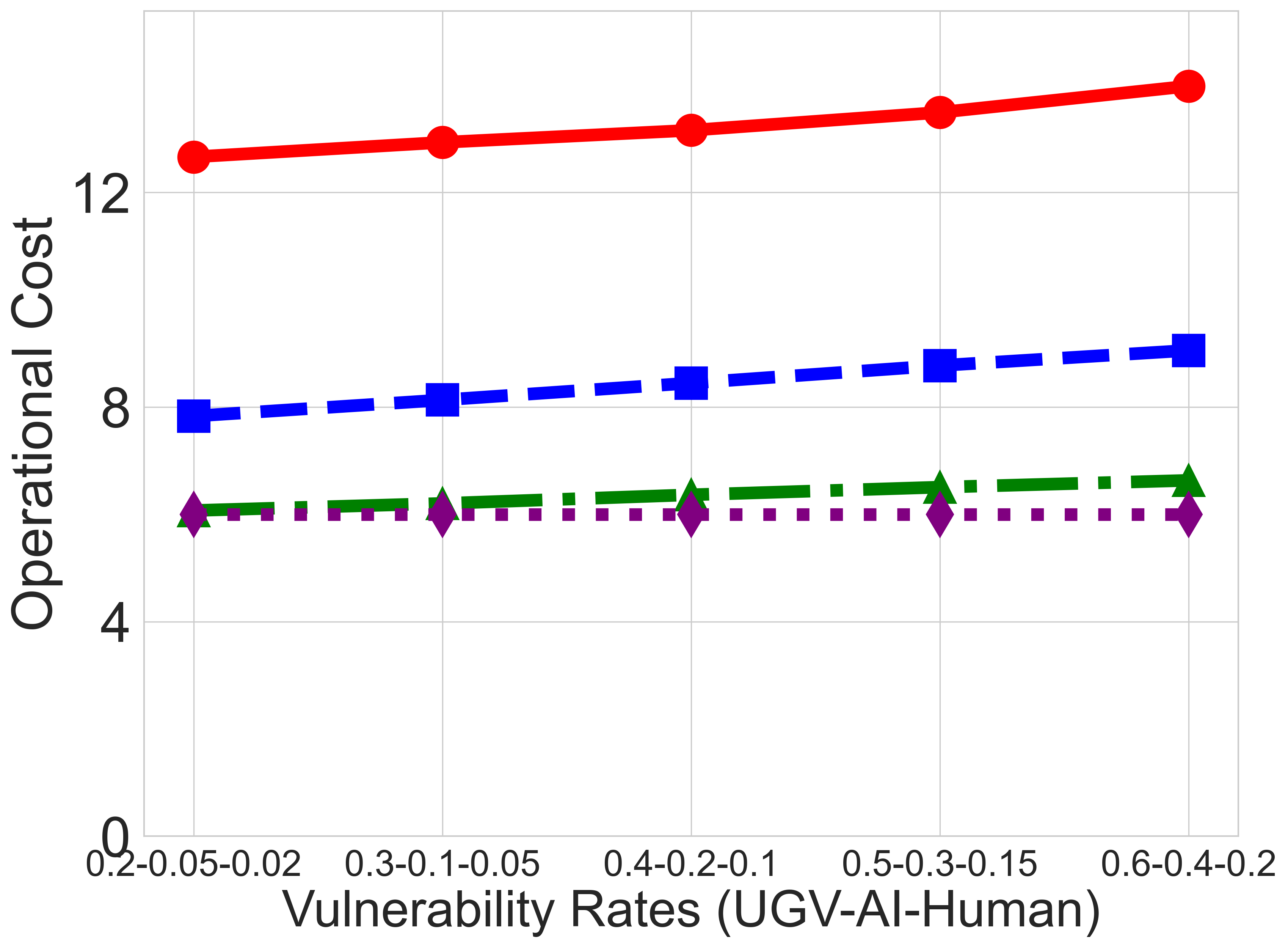}}
\hfil
\caption{Operational cost analysis of four schemes under varying attack rates (0.0 to 1.0).}
\label{fig:vulnerability_sensitivity_operational_cost}
\vspace{-3mm}
\end{figure*}

\textbf{Fig.~\ref{fig:vulnerability_sensitivity_operational_cost}} shows DASH-DF incurs the highest operational costs due to frequent bait tasks and recovery, increasing gradually with vulnerability. Although DF-only and BASE incur lower costs, their effectiveness is poor. At peak vulnerability, DASH-DF’s $\sim$50\% higher cost yields a $\sim$400\% MSR improvement over BASE, justifying the security-performance tradeoff.

\subsubsection{Ratio of Compromised Members} \label{sec:vulnerability_sensitivity_compromised_members}
\begin{figure*}[t]
\centering
\subfloat{\includegraphics[width=0.5\textwidth]{figures/metrics_legend.png}}
\hfil

\vspace{-3mm}
\setcounter{subfigure}{0}
\subfloat[UGV Compromised Rate]{\includegraphics[width=0.25\textwidth]{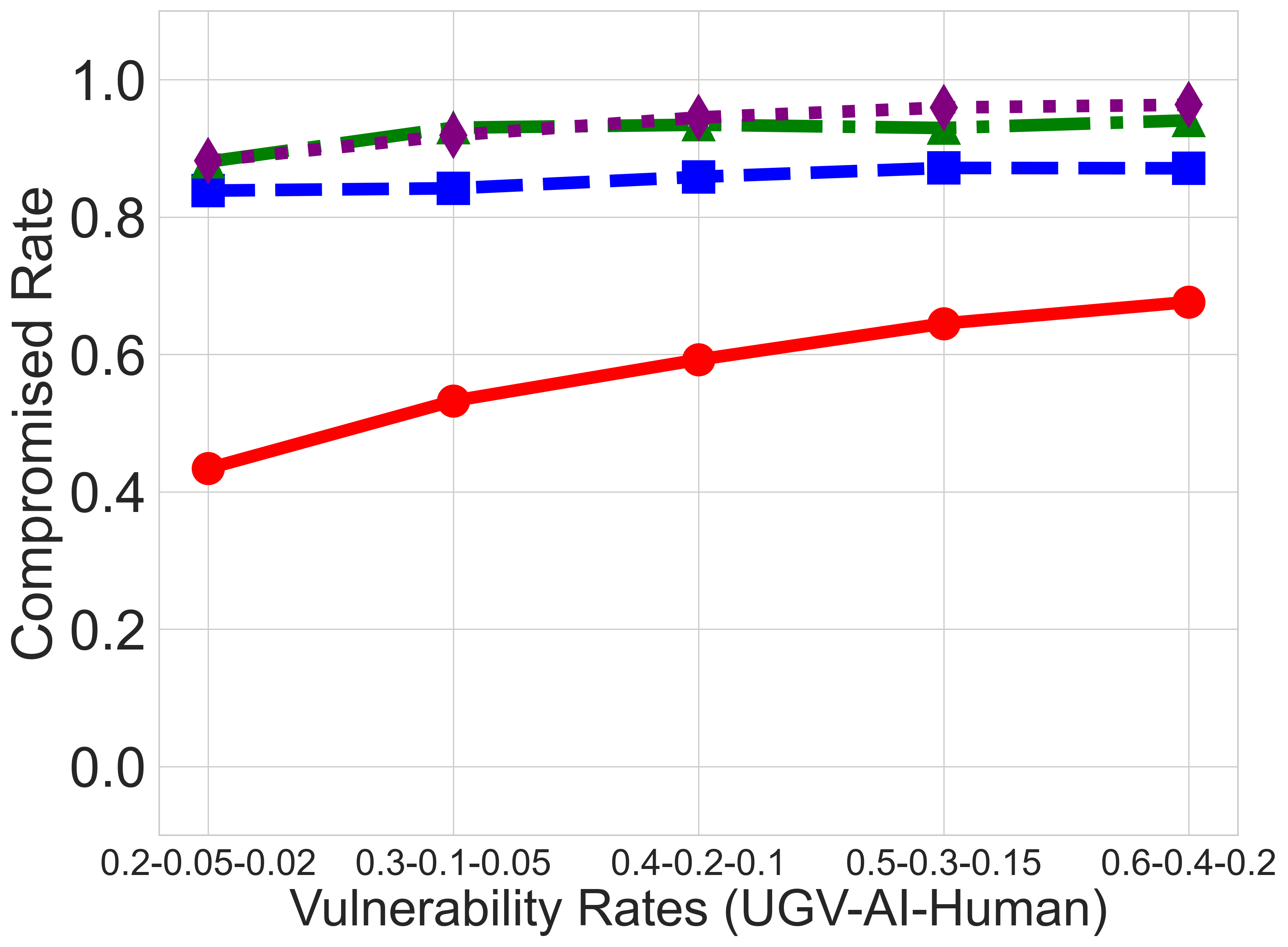}}
\hfil
\subfloat[AI Compromised Rate]{\includegraphics[width=0.25\textwidth]{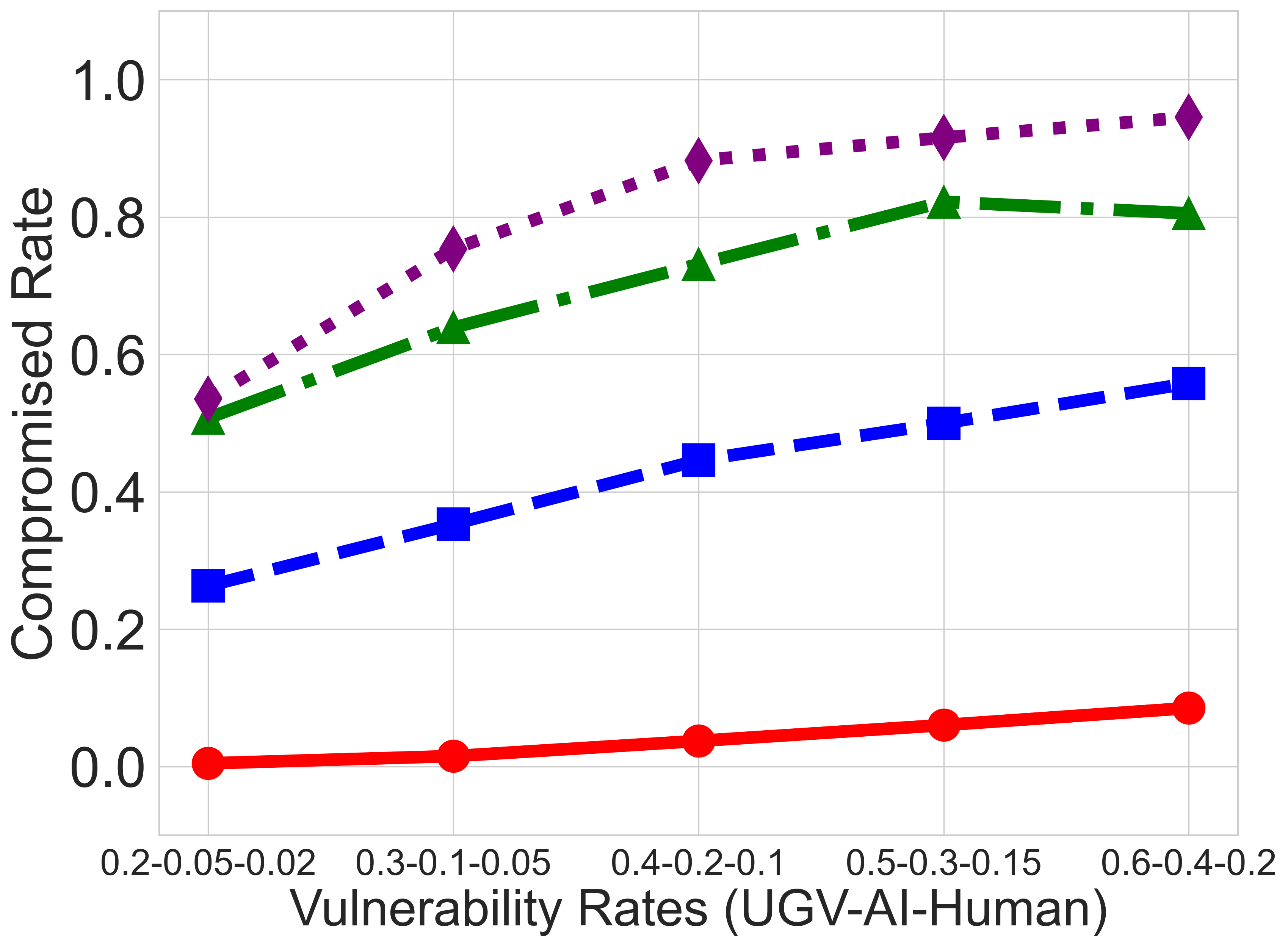}}
\hfil
\subfloat[Human Analysis Compromised Rate]{\includegraphics[width=0.25\textwidth]{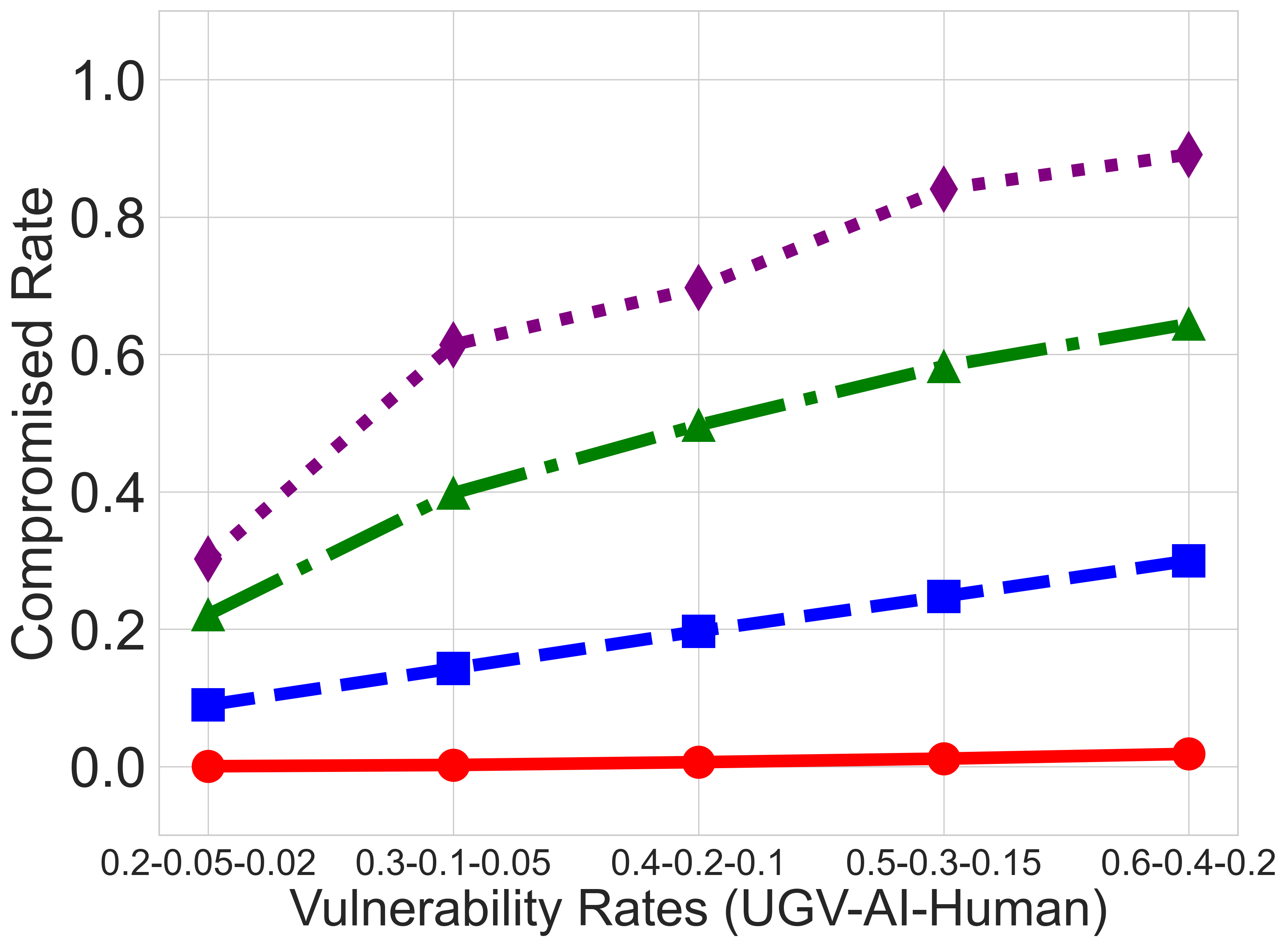}}
\hfil
\caption{Compromise rate analysis of four schemes under varying attack rates (0.0 to 1.0).}
\label{fig:vulnerability_sensitivity_compromised_members}
\vspace{-3mm}
\end{figure*}

\textbf{Fig.~\ref{fig:vulnerability_sensitivity_compromised_members}} shows compromise trends. DASH-DF limits UGV compromise to $\sim$0.65, well below other schemes (near 1.0). AI compromise in DASH-DF stays near 0.1, versus SMM-DF ($\sim$0.55), DF-only ($\sim$0.80), and BASE ($\sim$0.95). Human analysts are highly protected in DASH-DF (under 0.02), unlike BASE ($\sim$0.9), DF-only ($\sim$0.65), and SMM-DF ($\sim$0.3). These results confirm deception’s role in countering social engineering and protecting human assets.

\section{Conclusions \& Future Work} \label{sec:conclusion-and-future-work}

Our research advances Human-Machine Teaming Systems with the Deception-Augmented Shared Mental Model for Human-Machine Teaming (DASH) framework, which integrates security into coordination, an aspect overlooked in traditional approaches. DASH introduces a practical implementation of Shared Mental Models for multi-UGV coordination, moving beyond theoretical formulations. It also integrates component-specific deception via strategically deployed ``bait tasks'' and demonstrates how this approach enhances both mission performance and security. By bridging collaborative efficiency with system integrity, DASH establishes a new paradigm for resilient human-machine teaming in adversarial settings, with applications extending beyond surveillance to any secure multi-agent coordination task.

We found the following \textbf{key findings} from this study.  The DASH framework sustained $\simeq$60\% mission success at the highest attack frequency (1.0), compared to the baseline system's collapse below 10\%, demonstrating a sixfold improvement in resilience. Its SMM Quality Index consistently outperformed other schemes, maintaining $\simeq$60\% information accuracy even under maximum attack rates, confirming deception’s critical role in preserving information integrity. The SMM Coverage Index remained stable at $\simeq$0.8 across all attack rates, ensuring effective communication despite adversarial pressure. DASH-DF significantly reduced compromise rates, keeping human analyst compromises below 0.1 across all attack frequencies, whereas other schemes approached 0.8, showcasing its superior protection. Although DASH-DF incurred 20–30\% higher operational costs than other schemes, this investment translated directly into improved mission success and dramatically reduced compromise rates, demonstrating a favorable security-performance tradeoff.

\textbf{Future work} will focus on several key directions. First, we will develop adaptive adversarial models that evolve alongside our defense mechanisms to enable more rigorous testing. Second, we will deploy the framework on physical testbeds with real UGVs, AI systems, and human operators to validate the simulation results. Third, we will enhance deception mechanisms by optimizing a library of component-specific bait tasks using reinforcement learning. For example, planned experiments include urban patrol trials to study robot-human collaboration in city environments and forest fire reconnaissance missions to assess SMM integrity under conditions of smoke and occlusion. Additionally, we will explore the ethical implications of deception-based security in human–machine teams, particularly focusing on its impact on trust dynamics and transparency requirements.

\section*{Acknowledgment}
This work is partly supported by the Army Research Office and Army Research Laboratory under Grant Contract Numbers W911NF-24-2-0241 and NSF award 2107450. The views and conclusions contained in this document are those of the authors. They should not be interpreted as representing the official policies, expressed or implied, of the Army Research Laboratory or the U.S. Government. The U.S. Government is authorized to reproduce and distribute reprints for Government purposes, notwithstanding any copyright notation herein.

\bibliographystyle{IEEEtranN}
\bibliography{ref}

\end{document}


\title{DASH: Deception-Augmented Shared Mental Model for a Human-Machine Teaming System: Supplement}

\author{Zelin Wan, Han Jun Yoon, Nithin Alluru, Terrence J. Moore, Frederica F. Nelson, Seunghyun Yoon, Hyuk Lim, Dan Dongseong Kim, and Jin-Hee Cho,~\IEEEmembership{Senior Member, IEEE}
\IEEEcompsocitemizethanks{\IEEEcompsocthanksitem This research is partially supported by the DEVCOM ARL Army Research Office (ARO) Award (W911NF-24-2-0241), the National Science Foundation (NSF) Secure and Trustworthy Cyberspace (SaTC) Award (2330940), and the Virtual Institutes for Cyber and Electromagnetic Spectrum Research and Employment (VICEROY) program under the Air Force Research Laboratory (AFRL) initiatives through The Griffiss Institute (419890). The views and conclusions contained in this document are those of the authors and should not be interpreted as representing the official policies, either expressed or implied, of the Army Research Laboratory or the U.S. Government. The U.S. Government is authorized to reproduce and distribute reprints for Government purposes, notwithstanding any copyright notation herein. ({\em Corresponding author: Zelin Wan}).  Zelin Wan, Nithin Alluru, and Jin-Hee Cho are with the Department of Computer Science, Virginia Tech, Arlington, VA, USA. Email: \{zelin, nithin, jicho\}@vt.edu.  Terrence J. Moore and Frederica F. Nelson are with the US Army DEVCOM Army Research Laboratory, Adelphi, MD, USA. Email: \{terrence.j.moore.civ, frederica.f.nelson.civ\}@army.mil.  Sunghyun Yoon and Hyuk Lim are with the Korea Institute of Energy Technology (KENTECH), Naju-si, Jeollanam-do, Republic of Korea. Email: \{syoon, hlim\}@kentech.ac.kr.  Dan Dongseong Kim is with the University of Queensland, Brisbane, Queensland, Australia. Email: dan.kim@uq.edu.au.
}}

\markboth{IEEE Transactions on Human Machine Systems,~Vol.~xx, No.~x, xxx~2025}%
{Wan \MakeLowercase{\textit{et al.}}: DASH: Deception-Augmented Shared Mental Model for a Human-Machine Teaming System}


\maketitle
\appendices

\begin{table*}[!ht]
\caption{Key Concepts and Components of the DASH Framework}
\label{table:smm_components}
\centering
\begin{tabular}{|P{0.2\linewidth}|p{0.7\linewidth}|}
\hline
\textbf{Component} & \multicolumn{1}{c|}{\textbf{Description}} \\
\hline
Shared Mental Model (SMM) & A collective knowledge structure derived from Individual Mental Models (IMMs), providing a unified understanding of tasks, roles, and situational awareness to enhance coordination and decision-making. \\
\hline
Individual Mental Models (IMMs) & Role-specific knowledge encapsulating mission goals, team roles, and resource requirements, forming the foundation for the SMM. \\
\hline
Task Mental Model (TAMM) & Part of the IMM, encompassing information about mission goals (e.g., target types for object detection), expected outcomes (e.g., images from UGVs, predictions from AI agent), and task requirements. Facilitates effective execution and collaboration by clarifying each member's responsibilities. \\
\hline
Team Mental Model (TEMM) & Part of the IMM, including perceptions of team members' roles (e.g., sensing, identification), capabilities, and equipment. Promotes effective collaboration by ensuring tasks are allocated based on each member's strengths. \\
\hline
Adaptive Deceptive Task Management (ADTM) & A mechanism for task creation and allocation that integrates deception techniques to proactively detect and mitigate compromised agents. ADTM dynamically assigns both regular mission tasks and bait tasks to team members to maintain mission progress while ensuring system integrity. \\
\hline
Regular Tasks & Tasks designed to achieve primary mission objectives, such as data collection and analysis. Managed by ADTM to ensure mission progress. \\
\hline
Bait Tasks & Tasks introduced to identify potentially compromised agents by testing their integrity through predetermined tasks with known outcomes. Includes targeted verification tasks tailored to UGVs, AI agents, and human analysts. \\
\hline
Help Requests & Requests generated by team members for assistance with high-uncertainty or challenging tasks, facilitating collaboration between AI agents and human analysts. \\
\hline
Uncertainty Evaluation & Assessment of prediction uncertainty ($u$) by the AI agent, triggering validation or assistance requests when thresholds ($u > \tau_u$) are exceeded. \\
\hline
Detection Threshold & Predefined thresholds (e.g., $\zeta$ for trust) used to initiate additional validation steps or detect anomalies, such as triggering bait tasks when trust falls below acceptable levels. \\
\hline
Integrity Testing & Verification of team members' trustworthiness through tasks designed to detect anomalies or compromises, utilizing bait tasks to assess performance integrity. \\
\hline
Trust Management & Mechanisms to monitor, adjust, and evaluate trust levels of team members based on their behavior and task performance, including dynamic trust calculations using the number of correctly handled tasks. \\
\hline
Compromise Indicators & Signs of potential compromise, such as discrepancies in task outputs or elevated uncertainty levels, used within integrity testing and trust management processes. \\
\hline
Compromise Recovery & Recovery mechanisms, such as system reinstallation for UGVs, retraining for AI agents, and personnel replacement for human analysts, ensuring system integrity after detecting compromised agents. \\
\hline
Team Member's Skill & Awareness of specific capabilities and limitations of team members, informing task allocation to maximize effectiveness. \\
\hline
Team Member's Needs & Anticipation of informational and support requirements, ensuring timely resource allocation and assistance. \\
\hline
Team Member's State & Real-time operational status (e.g., active, idle, compromised), allowing dynamic task reallocation and intervention when necessary. \\
\hline
\end{tabular}
\end{table*}

\begin{table*}[t]
\caption{\sc Mathematical Notations}
\label{table:notation}
\centering
\begin{tabular}{|c|p{0.7\linewidth}|}
\hline
\textbf{Notation} & \multicolumn{1}{c|}{\textbf{Meaning}} \\
\hline
$C_i$ & Binary indicator for mission $i$ success (1 if successful, 0 otherwise) \\
\hline
MSR & Mission Success Rate (proportion of successful missions) \\
\hline
$C_{\text{total}}$ & Total operational cost across all missions \\
\hline
$T_i$ & Time to complete mission $i$ \\
\hline
$T_{\text{max}}$ & Maximum allowed operation time \\
\hline
$\omega_1$, $\omega_2$ & Optimization weighting factors (balancing success vs. cost) \\
\hline
$T_X (t)$ & Trust level of member $X$ (UGV, AI agent, or human analyst) at time $t$ \\
\hline
$N_C$ & Number of tasks where output matched consensus \\
\hline
$N_{\text{total}}$ & Total number of assigned tasks \\
\hline
$\gamma$ & Trust decay rate parameter \\
\hline
$t$ & Elapsed mission time \\
\hline
$\lambda$ & Trust initialization constant \\
\hline
$\zeta$ & Trust threshold for triggering bait task verification \\
\hline
$\tau_u$ & AI uncertainty threshold for human involvement \\
\hline
$R_X$ & Ratio of compromised status for member $X$ \\
\hline
SQI & Shared Mental Model Quality Index (accuracy of shared information) \\
\hline
SCI & Shared Mental Model Coverage Index (completeness of information sharing) \\
\hline
$I_{\text{correct}}$ & Number of correctly shared information pieces \\
\hline
$I_{\text{shared}}$ & Number of actually exchanged information pieces \\
\hline
$I_{\mathrm{max}}$ & Maximum number of shareable information pieces \\
\hline
$u$ & AI epistemic uncertainty (vacuity) due to lack of evidence \\
\hline
$\mathbf{b}$ & AI belief vector (per-class belief values) \\
\hline
$\mathbf{a}$ & Human analyst base rate vector (prior probabilities) \\
\hline
$\mathbf{P}_x$ & Expected belief of state $x$ \\
\hline
\end{tabular}
\end{table*}

\begin{figure*}
    \centering
    \includegraphics[width=\textwidth]{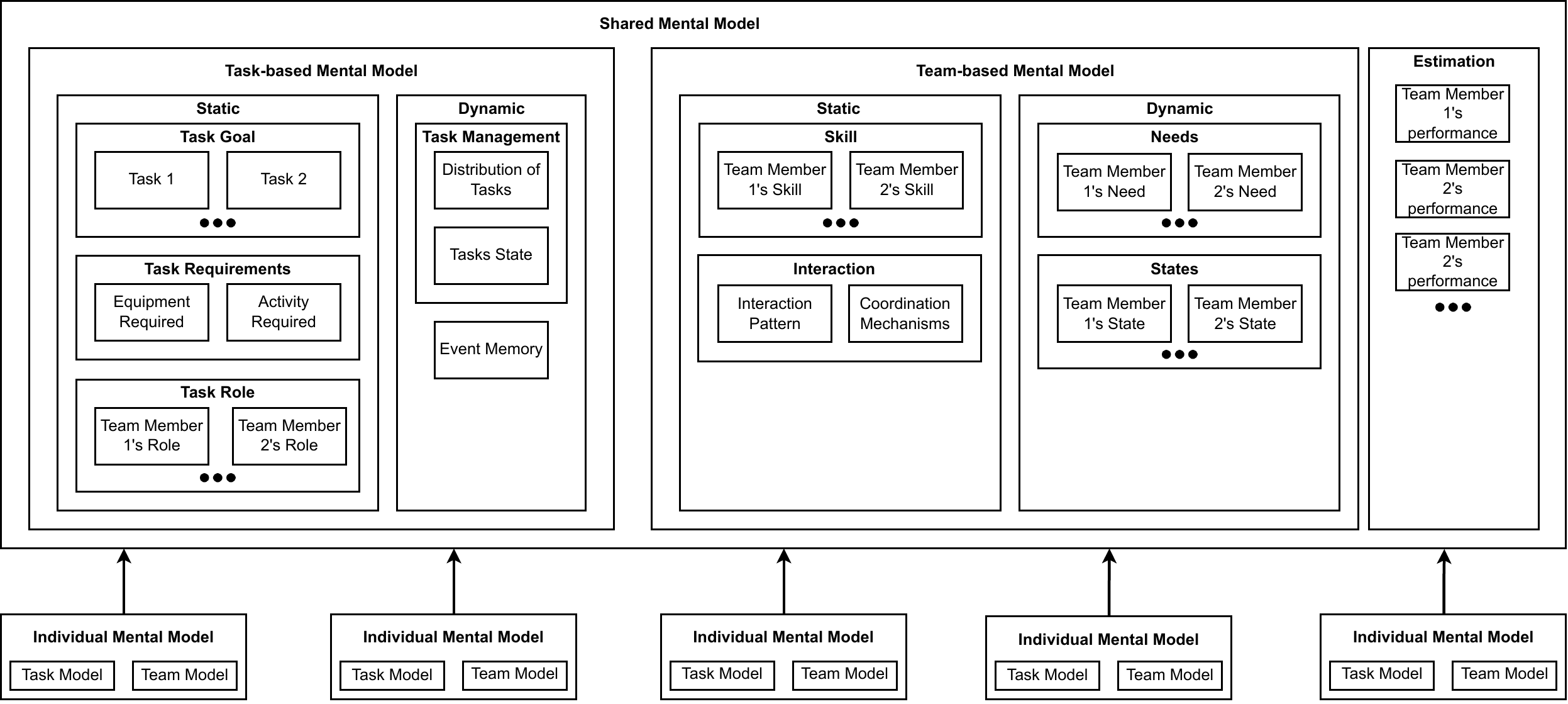}
    \caption{\sc The Proposed Shared Mental Model (SMM) where each component is explained in Table~\ref{table:smm_components}.}
    \label{fig:smm_framework}
\end{figure*}

\begin{figure*}
    \centering
    \includegraphics[width=\textwidth]{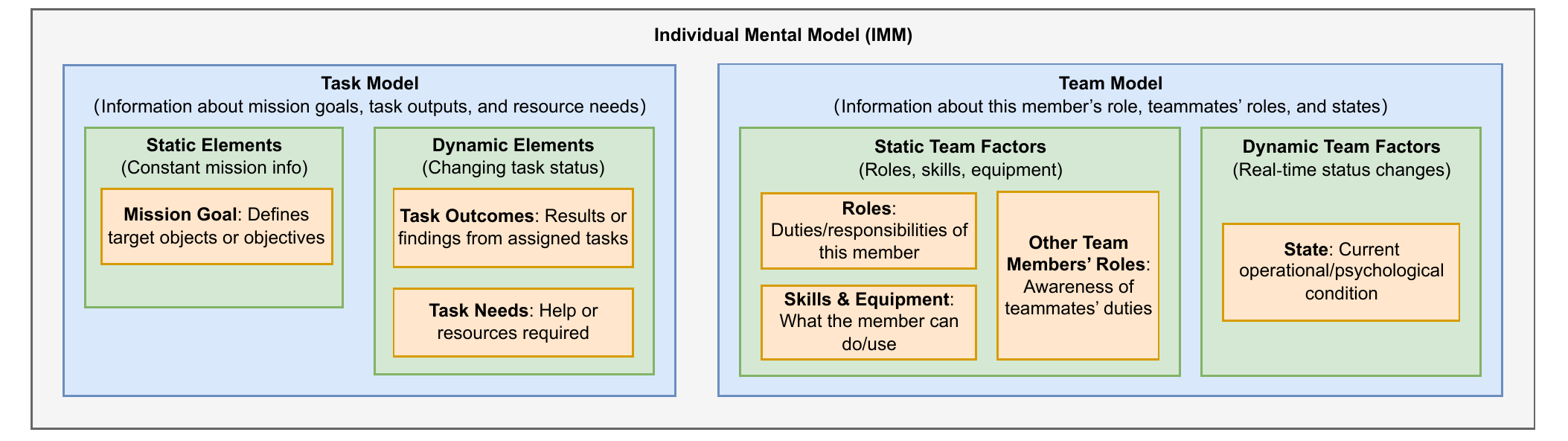}
    \caption{Individual Mental Model (IMM) framework where each component is detailed in Table~\ref{table:smm_components}.}
    \label{fig:imm_framework}
\end{figure*}

\begin{table*}[h]
\caption{\sc Experimental Parameters, Their Meanings, and Default Values}
\label{table:exp_parameters}
\centering
\begin{tabular}{|P{1.5cm}|p{8cm}|P{4cm}|}
\hline
\textbf{Parameter} & \multicolumn{1}{|c|}{\textbf{Meaning}} & \textbf{Value} \\
\hline
$N_{\text{sim}}$ & Number of simulation runs per scheme to ensure statistical significance & 100 \\
\hline
$N_m$ & Number of missions per simulation & 200 \\
\hline
--- & Attack frequency increments (fraction of tasks affected by attacks) & 0.0, 0.1, 0.2, 0.3, 0.4, 0.5, 0.6, 0.7, 0.8, 0.9, 1.0 \\
\hline
--- & UGV attack success rate (probability of successful UGV compromise) & 30\% \\
\hline
--- & AI attack success rate (probability of successful AI compromise) & 10\% \\
\hline
--- & Human attack success rate (probability of successful human compromise) & 5\% \\
\hline
\multicolumn{3}{|c|}{\textbf{\gray Trust Management Parameters}} \\
\hline
$\zeta$ & Trust threshold for bait task triggering & 0.3 \\
\hline
$\lambda$ & Trust initialization parameter ensuring initial trust level = 1 & 10 \\
\hline
$\gamma$ & Trust decay parameter controlling trust reduction over time & 0.001 \\
\hline
\multicolumn{3}{|c|}{\textbf{\gray Operational Cost Parameters}} \\
\hline
$C^A_{\mathrm{detect}}$ & AI agent computational cost per detection & 0.1 \\
\hline
$C^D_{\mathrm{detect}}$ & Cost when human analyst assists with verification & 1.0 \\
\hline
\multicolumn{3}{|c|}{\textbf{\gray Recovery \& Downtime Parameters}} \\
\hline
$DT_{\text{UGV}}$ & UGV reinstallation delay (in mission times) & 2 \\
\hline
$DT_{\text{AI}}$ & AI agent retraining delay (in mission times) & 3 \\
\hline
$DT_{\text{human}}$ & Human analyst replacement delay (in mission times) & 5 \\
\hline
\multicolumn{3}{|c|}{\textbf{\gray Operational Parameters}} \\
\hline
$\tau_u$ & Uncertainty threshold for an AI agent to request human help & 0.25 \\
\hline
--- & Number of standby UGVs for backup & 5 \\
\hline
--- & Number of standby AI agents for backup & 3 \\
\hline
--- & Number of object types for detection & 5 \\
\hline
--- & Required target detections per mission & 10 \\
\hline
\multicolumn{3}{|c|}{\textbf{\gray AI Model Training Parameters}} \\
\hline
--- & Dataset size for AI agent training & 6,000 surveillance images \\
\hline
--- & Optimizer & Adam \\
\hline
--- & Learning rate & 0.0001 \\
\hline
--- & Batch size & 32 \\
\hline
--- & Number of training epochs & 10,000 \\
\hline
\multicolumn{3}{|c|}{\textbf{\gray Sensitivity Analyses Parameters}} \\
\hline
--- & Vulnerability triplets (UGV:AI:Human) & \{0.2:0.05:0.02\}, \{0.3:0.1:0.05\}, \{0.4:0.2:0.1\}, \{0.5:0.3:0.15\}, \{0.6:0.4:0.2\} \\
\hline
$\omega_1, \omega_2$ & Weighting factors for mission success vs. cost optimization & Not specified (with constraint $\omega_1 + \omega_2 = 1$) \\
\hline
\end{tabular}
\end{table*}

\section{Detailed Trust Management Algorithms}
This section provides detailed pseudocode for the trust management algorithms implemented in the DASH framework. These algorithms are critical components of the framework's security mechanisms, enabling dynamic trust assessment and information sharing among team members.

\subsection{Trust Parameter Updates}
Algorithm~\ref{algorithm:N-conflict-free} details the procedure for updating trust parameters based on task outcomes. When AI and human analyst results match, the system increments the credible task counter, indicating alignment between team members. This algorithm directly supports the trust calculation formula in Eq.~\eqref{eq:trust} from the main paper, where trust is computed based on the ratio of consistent results to total tasks, with time-based decay.

\begin{equation} \label{eq:trust}
T_X (t) = \frac{N_{C}+\lambda}{N_{\mathrm{total}}+\lambda} \cdot e^{-\gamma \cdot t}, \quad T_X (t) \in [0,1].
\end{equation}

\begin{algorithm}
\caption{Trust Parameter Updates for Eq.~\eqref{eq:trust}}
\label{algorithm:N-conflict-free}
\begin{algorithmic}[1]
\Require Initial values: 
\State $N_{C} \gets 0$ \Comment{Number of credible tasks}
\State $N_{\mathrm{total}} \gets 0$ \Comment{Total number of tasks assigned}
\State $t \gets 0$ \Comment{Execution time of a given mission}

\Procedure{UpdateTrustMetrics}{task}
    \If{AI\_result = Human\_analyst\_result}
        \State $N_{C} \gets N_{C} + 1$ \Comment{Increment if results match}
    \EndIf
    \State $N_{\mathrm{total}} \gets N_{\mathrm{total}} + 1$
    \State $t \gets t + 1$
    
    \If{BaitTaskPassed} \Comment{Reset trust after successful verification}
        \State $N_{C} \gets 0$, $N_{\mathrm{total}} \gets 0$, $t \gets 0$
    \EndIf
\EndProcedure
\end{algorithmic}
\end{algorithm}

\subsection{Trust-based Information Sharing}
Algorithm~\ref{algorithm:trust_proportional_exchange} implements the probabilistic information sharing mechanism that proportionally restricts information flow based on current trust levels. This approach ensures that information is shared with team members in proportion to their trustworthiness, reducing the risk of critical information being compromised. The algorithm uses the trust value calculated by Eq.~\eqref{eq:trust} to determine whether information should be shared in any given instance.

\begin{algorithm}
\caption{Trust-based Information Sharing}
\label{algorithm:trust_proportional_exchange}
\begin{algorithmic}[1]
\Require Information types: sensed images, UGV locations, status, AI queue order, detection results  
\Procedure{AttemptInfoSharing}{$X$, $Y$, $info$}
    \State $T_Y (t) \gets$ Eq.~\eqref{eq:trust} \Comment{Retrieve $Y$'s trust at time $t$}
    \State $r \gets \mathrm{random}(0, 1)$ \Comment{Generate random value}
    
    \If{$r \leq T_Y (t)$}
        \State $\mathrm{ShareInfo} (X, Y, info)$ \Comment{Transmit information}
        \State \Return TRUE
    \Else
        \State $\mathrm{LogBlockedTransmission}(X, Y, info)$ \Comment{Block transmission}
        \State \Return FALSE
    \EndIf
\EndProcedure
\end{algorithmic}
\end{algorithm}
